

\input epsf     


\magnification=\magstephalf


\def\etal{{\it et~al.\ }}
\def\la{\langle}
\def\ra{\rangle}
\def\tm#1{\widetilde M_{#1}}
\def\ts#1{\widetilde S_{#1}}
\def\dd{\langle\delta|\delta|\rangle}
\def\dmod{\langle|\delta|\rangle}
\def\dmodr{\langle|\delta|\rangle-(2/\pi)^{1/2}\sigma}
\def\dsqr{\langle\delta^2\rangle}
\def\dcube{\langle\delta^3\rangle}
\def\dfour{\langle\delta^4\rangle}
\def\tt{\langle\theta|\theta|\rangle}

\def\tsqr{\langle\theta^2\rangle}
\def\tcube{\langle\theta^3\rangle}
\def\tfour{\langle\theta^4\rangle}
\def\gg{\langle\delta_g|\delta_g|\rangle}

\def\gcube{\langle\delta_g^3\rangle}

\def\nbody{$N$-body}
\def\e{\varepsilon}
\def\k{\kappa}
\def\hmpc{h^{-1}\,{\hbox{Mpc}}}
\newbox\grsign \setbox\grsign=\hbox{$>$} \newdimen\grdimen \grdimen=\ht\grsign
\newbox\simlessbox \newbox\simgreatbox
\setbox\simgreatbox=\hbox{\raise.5ex\hbox{$>$}\llap
     {\lower.5ex\hbox{$\sim$}}}\ht1=\grdimen\dp1=0pt
\setbox\simlessbox=\hbox{\raise.5ex\hbox{$<$}\llap
     {\lower.5ex\hbox{$\sim$}}}\ht2=\grdimen\dp2=0pt
\def\simgt{\mathrel{\copy\simgreatbox}}
\def\simlt{\mathrel{\copy\simlessbox}}

\def\chaphead{}
\newcount\eqnumber
\eqnumber=1
\def\new{{\rm\chaphead\the\eqnumber}\global\advance\eqnumber by 1}
\def\eqref#1{\advance\eqnumber by -#1 \chaphead\the\eqnumber
     \advance\eqnumber by #1 }
\def\last{\advance\eqnumber by -1 {\rm\chaphead\the\eqnumber}\advance
     \eqnumber by 1}

\def\capskip{\baselineskip = 10 pt}
\def\textskip{\baselineskip = 12pt plus 1pt minus 1pt}
   \font\eightrm=cmr8   \font\sixrm=cmr6
   \font\eighti=cmmi8   \font\sixi=cmmi6
  \font\eightsy=cmsy8  \font\sixsy=cmsy6
  \font\eightbf=cmbx8  \font\sixbf=cmbx6
  \font\eighttt=cmtt8
  \font\eightit=cmti8
  \font\eightsl=cmsl8
\newskip\ttglue
\def\eightpoint{\def\rm{\fam0\eightrm}
  \textfont0=\eightrm \scriptfont0=\sixrm \scriptscriptfont0=\fiverm
  \textfont1=\eighti  \scriptfont1=\sixi  \scriptscriptfont1=\fivei
  \textfont2=\eightsy \scriptfont2=\sixsy \scriptscriptfont2=\fivesy
  \textfont3=\tenex \scriptfont3=\tenex \scriptscriptfont3=\tenex
  \textfont\itfam=\eightit  \def\it{\fam\itfam\eightit}%
  \textfont\slfam=\eightsl  \def\sl{\fam\slfam\eightsl}%
  \textfont\ttfam=\eighttt  \def\tt{\fam\ttfam\eighttt}%
  \textfont\bffam=\eightbf  \scriptfont\bffam\sixbf
   \scriptscriptfont\bffam=\fivebf \def\bf{\fam\bffam\eightbf}%
  \tt \ttglue=0.5em plus .25em minus.15em
  \setbox\strutbox=\hbox{\vrule height7pt depth2pt width0pt}%
  \let\sc=\sixrm  \let\big=\eightbig  \normalbaselines\rm}


\textskip
\smallskip

\centerline{\bf WEAKLY NON-LINEAR GAUSSIAN FLUCTUATIONS}
\centerline{\bf AND THE EDGEWORTH EXPANSION}
\bigskip
\centerline{by Roman Juszkiewicz$^{1,2,3}$, David Weinberg$^3$, Piotr
Amsterdamski$^1$}
\centerline{Micha{\l} Chodorowski$^1$, and Fran{\c c}ois Bouchet$^{2,3}$}
\bigskip
\line{$^1$Copernicus Astronomical Center, ul. Bartycka 18, PL-00-716 Warszawa,
Poland\hfil}
\line{$^2$Institut d'Astrophysique de Paris, CNRS, 98 bis Blvd. Arago,
F-75015 Paris, France\hfill}
\line{$^3$Institute for Advanced Study, Princeton, NJ 08540, USA\hfill}

\bigskip
\line{E-mail addresses: roman@camk.edu.pl, dhw@guinness.ias.edu,
pa@camk.edu.pl,
\hfil}
\line{\phantom{E-mail addresses:} michal@camk.edu.pl, bouchet@iap.fr\hfil}

\vskip 0.25truein

\centerline{\bf ABSTRACT}
\medskip

We calculate the cosmological evolution of the 1-point probability
distribution function (PDF), using an analytic approximation that combines
gravitational perturbation theory with the Edgeworth expansion of the PDF.
Our method applies directly to a smoothed mass density field or to the
divergence of a smoothed peculiar velocity field, provided that r.m.s.\
fluctuations are small compared to unity on the smoothing scale, and that
the primordial fluctuations that seed the growth of structure are Gaussian.
We use this ``Edgeworth approximation'' to compute the evolution of
$\dd$ and $\dmod$; these measures are similar to the skewness and kurtosis
of the density field, but they are less sensitive to the tails of the
probability distribution, so they may be more accurately estimated from
surveys of limited volume.  We compare our analytic calculations to the
results of cosmological \nbody\ simulations in order to assess their range
of validity. The Edgeworth
approximation for the PDF, and the computations of $\dd$ and $\dmod$
that are based on it, remain quite accurate until the r.m.s.\
density fluctuation
is $\sigma \sim 1/2$, or, more generally, until the magnitude of the
skewness approaches one.  The skewness and kurtosis of the density field
stay remarkably close to the values predicted by perturbation theory
even when the r.m.s.\  fluctuation is $\sigma=2$.
When $\sigma \ll 1$, the numerical simulations and perturbation theory
agree precisely, demonstrating that the \nbody\ method can yield accurate
results in the regime of weakly non-linear clustering.
We show analytically that
``biased'' galaxy formation preserves the relation $\dcube \propto \dsqr^2$
predicted by second-order perturbation theory, provided that the galaxy
density is a local function of the underlying mass density.
The constant of proportionality depends on the shape of the
biasing function that relates galaxy and mass densities.
Our results should be useful in the analysis of large-scale density and
velocity fields, allowing one to derive constraints on the nature of
primordial fluctuations, the value of the cosmological density parameter,
and the physical processes that govern galaxy formation.

\medskip
\line{{\it Subject Headings:} cosmology: theory --- large-scale structure of
the Universe \hfil}

\vfill\eject
\parskip=8pt

\noindent
{\bf 1. Introduction}

Most theories for the formation of structure in the universe assume that
this structure developed by gravitational instability from small-amplitude
primordial fluctuations.  The simplest hypothesis is that these fluctuations
were Gaussian, and simple versions of inflationary cosmology naturally
produce Gaussian fluctuations from the quantum fluctuations of the inflaton
field (Guth \& Pi 1982; Hawking 1982; Starobinsky 1982; Bardeen, Steinhardt \&
Turner 1983).  Topological defect models predict non-Gaussian fluctuations, as
do some specialized versions of inflation involving multiple scalar fields
(e.g.\ Zel'dovich 1980; Turok 1989; Barriola \& Vilenkin 1989;
Allen, Grinstein \& Wise 1987; Kofman \& Pogosyan 1988; Salopek, Bond \&
Bardeen 1989).
Observational evidence for or against Gaussian primordial fluctuations
can therefore provide important clues to physics in the very early universe,
and to the physical origin of today's large-scale structure.

In the linear approximation for gravitational instability, fluctuations that
are initially Gaussian remain Gaussian.  However, non-linear effects
quickly distort Gaussian fluctuations, and they are quite significant on all
scales that can be probed by current observational surveys.
In this paper we show how to compute the evolution of the 1-point probability
distribution function (PDF) in the weakly non-linear regime, incorporating
systematically the
first departures from Gaussianity.  We also compute
the evolution of two quantities that measure the asymmetry and width of the
PDF, respectively $\dd$ and $\dmod$,
where $\delta\equiv(\rho-\la\rho\ra)/\la\rho\ra$
is the density contrast and the brackets $\la \ldots \ra$ denote averaging
over the PDF.
These measures are similar to the skewness
and kurtosis of the density field,
but they may be less subject to
observational sampling errors because they are
less dependent on the high-$\delta$ tail of the PDF.
We present similar calculations for the divergence of the
velocity field, and we consider the effects of ``biased'' galaxy formation
on the moments and PDF of the density field.
Throughout the paper we compare our analytic calculations to cosmological
$N$-body simulations in order to assess their range of validity.
Our results should be useful in the analysis of large-scale density and
velocity fields, providing tools with which to test the hypothesis of
Gaussian initial fluctuations and constrain the value of $\Omega$, the
cosmological density parameter.

There have been previous efforts to compute the evolution of the PDF
from Gaussian initial conditions, employing the Zel'dovich approximation
(Kofman \etal 1993) as well as rigorous perturbation theory (Bernardeau
1992).  However,
these methods compute the probability distribution of the {\it unsmoothed}
final density field that evolves from {\it smoothed} initial conditions.
For comparison to observational data, the relevant quantity is the PDF of the
smoothed final density field that evolves from unsmoothed initial conditions,
which may be far into the non-linear regime on scales below the smoothing
length.  Because dynamical evolution is non-linear, the effects of gravity
and smoothing do not commute.
Padmanabhan \& Subramanian (1993), recognizing this problem, have
attempted to compute the PDF of the smoothed final density field
via the Zel'dovich approximation, i.e.\ by using first-order perturbation
theory in Lagrangian space.
In a series of recent papers, we have
shown how to calculate moments of the PDF of a smoothed final density field
using perturbation theory;
a rigorous perturbative calculation of order-$n$ moments requires
order-$(n-1)$ perturbation theory
(Juszkiewicz \& Bouchet 1992; Bouchet \etal 1992a;
Juszkiewicz, Bouchet \& Colombi 1993; see also Goroff \etal 1986).  In this
paper we combine
these results with the Edgeworth expansion (see Cram{\'e}r 1946) to obtain an
approximation to the full PDF.

\bigskip

\noindent
{\bf 2. The Edgeworth Expansion}

We wish to examine how
gravitational instability drives the PDF
away from its initial state, which we assume to
be Gaussian. We first introduce the Gram-Charlier expansion, which
allows one to reconstruct the PDF from its moments.
We then summarize the predictions of perturbation theory for the moments
of the smoothed mass density contrast $\delta$.
Finally, we rearrange the Gram-Charlier series by collecting
all terms of the same order. The result is the
proper asymptotic expansion of the PDF in powers of
$\sigma$, the standard deviation of $\delta$.

\topinsert
$$\vbox{\halign{\hfil$#$\tabskip=3em plus3em
minus3em&\hfil#\hfil&\hfil$#$\hfil\tabskip=1em&$#$\hfil\tabskip=3em
plus3em minus3em&\hfil #\hfil&\hfil#\hfil\tabskip=0pt\cr
\multispan6{\hfil\bf TABLE 1:\ The Hermite polynomials \hfil}\cr
\noalign{\bigskip\hrule\smallskip\hrule\smallskip}
{\ell}&\multispan2{\hfil $H_{\ell}(\nu)$ \hfil}\cr
\noalign{\smallskip\hrule\smallskip}
0&\multispan2{ $1$ \hfil}\cr
1&\multispan2{ $x$ \hfil}\cr
2&\multispan2{ $x^2-1$ \hfil}\cr
3&\multispan2{ $x^3-3x$ \hfil}\cr
4&\multispan2{ $x^4-6x^2+3$ \hfil}\cr
5&\multispan2{ $x^5-10x^3+15x$ \hfil}\cr
6&\multispan2{ $x^6-15x^4+45x^2-15$ \hfil}\cr
\noalign{\medskip\hrule\smallskip}
}}$$
\endinsert

Our object of study is $p(\nu)$, the PDF
of the density field in terms of the
standardized random variable $\nu \equiv
\delta /\sigma$.
The probability that the density contrast at a randomly chosen location
lies in the range $\nu < \delta/\sigma < \nu+d\nu$ is $p(\nu)d\nu$.
Let us also introduce
$\phi(\nu) = (2\pi)^{-1/2}\exp (-\nu^2/2)$, a Gaussian
(or Normal) PDF.
Since we want to describe evolution from
Gaussian initial conditions, it makes sense to consider an expansion
of $p(\nu)$ in terms of $\phi(\nu)$ and its derivatives.
The {\it Gram-Charlier series} (Cram{\'e}r 1946 and references therein)
provides such an expansion:
$$
p(\nu) = c_0\,\phi (\nu) + {c_1\over 1!}\,\phi^{(1)}(\nu)
+ {c_2\over 2!}\,\phi^{(2)}(\nu) + \ldots \; ,
\eqno(\new)
$$
where $c_\ell$ are constant coefficients. Superscripts
denote derivatives with respect to $\nu$:
$$
\phi^{(\ell)}(\nu) \equiv {d^{\ell}\phi \over d\nu^{\ell}} =
(-1)^{\ell}\, H_{\ell}(\nu)\,
\phi(\nu),
\eqno(\new)
$$
where $H_{\ell}$ is the Hermite polynomial of degree $\ell$.
Table 1 gives expressions for the first seven $H_{\ell}$.

The Hermite polynomials satisfy orthogonality relations (e.g. Abramowitz
\& Stegun 1964):
$$
\int_{-\infty}^{\infty} \; H_{\ell}(\nu)\;H_m(\nu)\;\phi(\nu)\;
d\nu = \cases{ 0, &if $\ell \ne m$ ; \cr
               \ell\,!  &otherwise. \cr}
\eqno(\new)
$$
Therefore, multiplying both sides of equation (1) by $H_\ell$
and integrating term by term yields
$$
c_{\ell} = (-1)^{\ell}\;\int_{- \infty}^{\infty}\; H_{\ell}(\nu)\;
p(\nu)\;d\nu \; .
\eqno(\new)
$$
Equation (4) gives $c_0 = 1, \; c_1 = c_2 = 0$, while for the
next four coefficients in the series we obtain
$$
c_{\ell} = (-1)^{\ell}\,S_{\ell}\sigma^{\ell-2} \, , \;\; {\rm for}\;
3\leq \ell \leq 5\; ; \quad c_6 = S_6 \sigma^4 + S_3^2 \sigma^2 \;.
\eqno(\new)
$$
Here
$$
S_{\ell} \equiv M_{\ell}\,/\,\sigma^{2\ell - 2} \; ,
\eqno(\new)
$$
with $M_{\ell}$ being the $\ell$:th {\it cumulant}\footnote{$^1$}{A cumulant
of arbitrary order $\ell$ is given by
$d^{\ell}\ln\langle e^{t\delta} \rangle \, /dt^{\ell}$,
evaluated at $t = 0$. Cumulants are often called
``reduced'' or ``connected'' moments.}
of $p(\delta)$.
Each cumulant
is a combination of the central moments
$$
\mu_{\ell} \equiv
\sigma^{\ell}\; \int_{-\infty}^{\infty}
\;p(\nu)\;\nu^{\ell}\;d\nu \; .
\eqno(\new)
$$
In particular,
$$\eqalign{
M_2 &= \mu_2=\sigma^2, \quad M_3=\mu_3, \quad M_4 = \mu_4 - 3\sigma^4, \cr
M_5 &= \mu_5 - 10\sigma^2 M_3, \quad M_6 = \mu_6 - 15\sigma^2 M_4 - 10M_3^2
- 15\sigma^6 \; .
}
\eqno(\new)
$$
For a zero-mean, Gaussian distribution, all cumulants vanish except $M_2$.
Ratios of the cumulants to the appropriate powers of $\sigma$
provide convenient measures of
deviations from the Gaussian shape.
For example, the {\it skewness} $M_3/\sigma^3$
measures the asymmetry of the distribution,
while the {\it kurtosis} $M_4/\sigma^4$
measures the flattening of the tails relative to a Gaussian.
The significance of the ``normalized cumulants'' $S_{\ell}$
will become clear shortly.

Thus far our discussion has been entirely general. Now let us consider
the application to the gravitational evolution of initially
Gaussian fluctuations. In the weakly non-linear regime, when the r.m.s.
amplitude of density fluctuations, $\sigma$, is smaller than unity,
the time evolution equations for the smoothed
density contrast can be solved approximately by using the
perturbative expansion
$$
\delta = \delta_1 + \delta_2 + \ldots .
\eqno(\new)
$$
Here $\delta_1 = O(\sigma)$ is
the solution of the equations with all non-linear terms set to zero;
$\delta_2 = O(\sigma^2)$ is the solution of equations of motion
with quadratic terms included iteratively by using $\delta_1$ as a
source, and so on (see, e.g., \S 18 in Peebles 1980). For an
Einstein-de Sitter ($\Omega =1$) cosmology, all terms in this
expansion are known (Fry 1984); for $\Omega \ne 1$, solutions
are available only for the first few terms (Bouchet \etal 1992a,
hereafter BJCP).
We assume that $\delta_1$ is a Gaussian field.
All of its statistical properties
are therefore determined by its power spectrum,
$P(k)\equiv\langle\delta_k^2\rangle$, where
$k$ is the comoving wavenumber.
Since one constructs the higher order
terms iteratively out of $\delta_1$, the initial power spectrum also
determines the non-linear, dynamically driven
deviations from Gaussian behavior.
One can compute the moments of the density field needed for the
Gram-Charlier expansion of $p(\nu)$ by raising both sides of
equation (9) to the appropriate power, then averaging.
The cumulants $M_{\ell}$ can then be expressed in terms of the $S_{\ell}$
parameters, multiplied by the appropriate powers of
$\sigma = \langle \delta_1^2 \rangle ^{1/2}$.
For details of the calculations, see
Goroff \etal (1986, hereafter GGRW)
and Juszkiewicz \etal (1993, hereafter JBC).
The lowest non-vanishing contribution to the skewness is
$$
M_3 = 3 \langle \delta_1^2\delta_2 \rangle = O(\sigma^4) \; ,
\eqno(\new)
$$
and the $\ell$:th cumulant is of order (Fry 1984; Bernardeau 1992)
$$
M_{\ell} = O(\sigma^{2\ell-2}) \; .
\eqno(\new)
$$
Equation (11) implies that, in perturbation theory, the normalized cumulants
$S_{\ell}$ defined by equation (6)
{\it are always of order unity}. In an Einstein-de Sitter universe, the
$S_{\ell}$ remain independent of time in the perturbative regime,
determined only by the slope of
the power spectrum near the smoothing scale (GGRW; JBC).
For $\Omega \ne 1$ models, only $S_3$ has been calculated.
Although
$S_3$ does change with time in this case,
the time dependence is extremely weak, and for identical spectra and
smoothing functions $S_3$ remains
within several per cent of the corresponding Einstein-de Sitter value
provided $0.05 \leq \Omega \leq 3$ (BJCP; JBC).

The normalized cumulants $S_\ell$ have both a dynamic and a static application:
they describe the time evolution of moments of the PDF at a fixed smoothing
scale, but they also describe the relation between moments of the PDF at
a fixed time on different smoothing scales.  In the latter case,
one must also include the scale-dependence of the $S_{\ell}$ if the
initial power spectrum is not scale-free.

\topinsert
\capskip
\centerline{
\epsfxsize=3.5truein
\epsfbox[110 432 459 720]{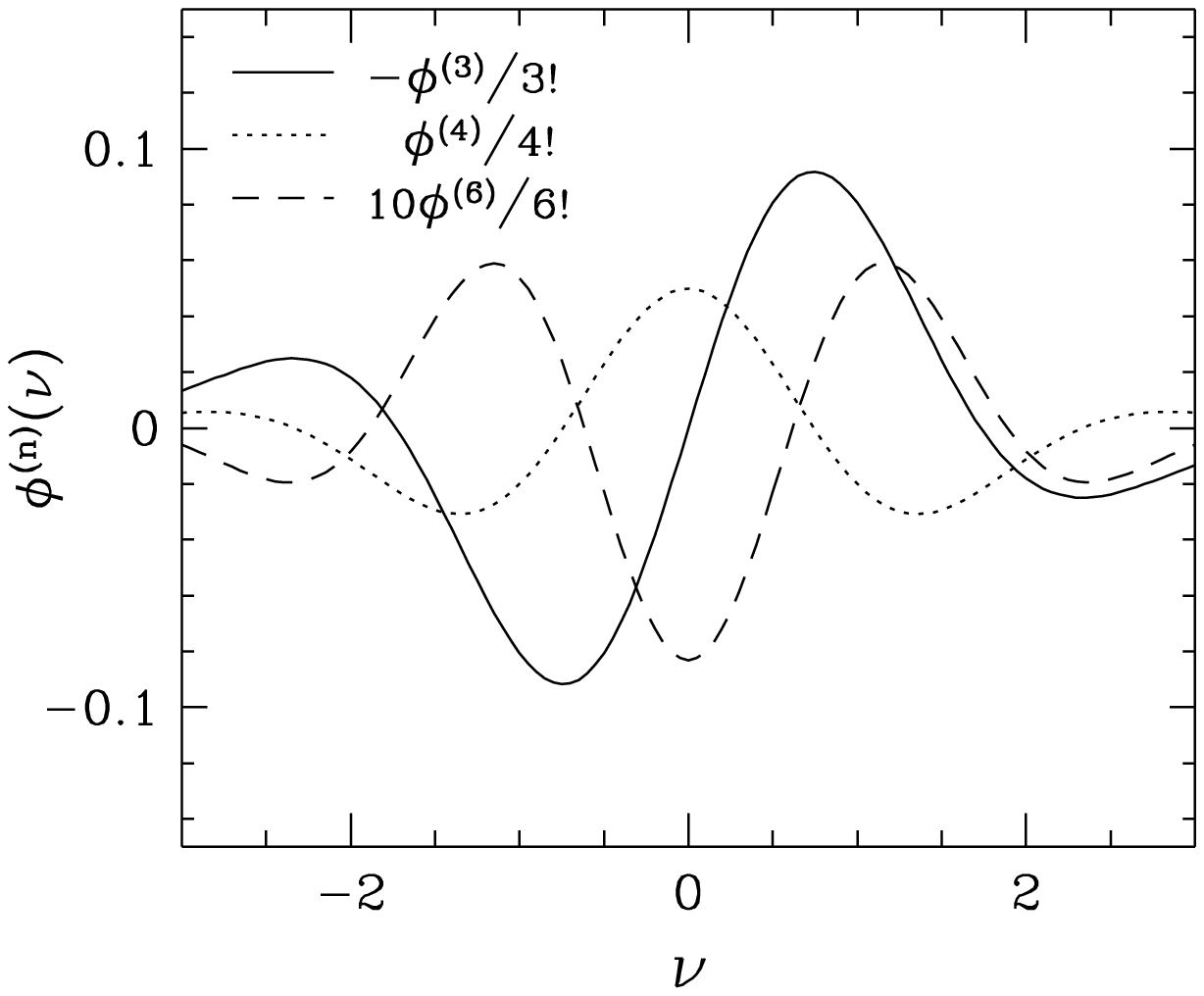}
}
\medskip
{\eightpoint
\noindent {\bf Figure 1} ---
The derivatives of the Gaussian function that appear
in the low-order terms of the Edgeworth series, equation (12).
Solid, dotted, and dashed lines show $\phi^{(3)}(\nu)$, $\phi^{(4)}(\nu)$,
and $\phi^{(6)}(\nu)$ respectively, multiplied by factors that appear
in equation (12).
}
\smallskip
\textskip
\endinsert

For our immediate purpose here, the important consequence of the fact
that the normalized cumulants are $O(1)$ in perturbation theory is
that the Gram-Charlier series {\it is not} a proper asymptotic expansion
for $p(\nu)$. In an asymptotic expansion,
the remainder term should be of higher order
than the last term retained.
However, if we truncated the
series (1) at the $\phi^{(4)}$ term, which is $O(\sigma^2)$,
we would miss another $O(\sigma^2)$ contribution coming from
$c_6$ (equation [5]). In order to deal with this problem, let
us rearrange the Gram-Charlier expansion
by collecting all terms with the same
powers of $\sigma$. The result is the so-called
{\it Edgeworth series}, with the first few terms given by
$$
p(\nu) = \phi(\nu) - {1\over 3!}S_3\phi^{(3)}(\nu)\sigma
+ {1\over 4!}S_4\phi^{(4)}(\nu)\sigma^2 +
{10\over 6!}S_3^2\phi^{(6)}(\nu)\sigma^2  + O(\sigma^3) \; .
\eqno(\new)
$$
Figure 1 plots the derivatives $\phi^{(3)}$, $\phi^{(4)}$, and
$\phi^{(6)}$ that appear in equation (\eqref{1}).
Cram{\'e}r (1946) lists the Edgeworth series to
higher order, and he proves that it is a proper asymptotic expansion.
This proof is directly relevant to our purposes, since
it implies that there are no additional $O(\sigma^2)$ terms hiding in the
Gram-Charlier series at $\ell > 6$.

Now we can see the attractiveness of the Edgeworth series for
describing the gravitational evolution of Gaussian fluctuations:
it becomes a series expansion for
the evolving PDF in powers of the r.m.s. fluctuation $\sigma$.
This makes physical sense because the Edgeworth series provides
an expansion about a Gaussian probability distribution.
If the initial fluctuations
are Gaussian, then we expect the terms describing successively larger
departures from a Gaussian PDF to come in with successively
higher powers of $\sigma$.
For similar reasons the Gauss-Hermite series, which is closely related
to the Edgeworth expansion, has recently found applications
in stellar dynamics as a description of galaxy line profiles
(e.g. van der Marel \& Franx 1993; Gerhard 1993).

Given equation (12), we can compute the Edgeworth approximation to the PDF
provided that we can compute $S_{\ell}$ to the required order.
In this paper we will make use of the second-order approximation,
$$
p(\nu) = \left[1+{1 \over 3!}S_3\sigma H_3(\nu)\right]\phi(\nu),
\eqno(\new)
$$
and the third-order approximation,
$$
p(\nu) =
\left[1+{1 \over 3!}S_3\sigma H_3(\nu) + {1\over 4!}S_4\sigma^2 H_4(\nu)
      + {10 \over 6!}S_3^2 \sigma^2 H_6(\nu) \right] \phi(\nu).
\eqno(\new)
$$
Although equation (13) contains only a single explicit power of $\sigma$,
it is appropriately described as a second-order approximation because
the parameter $S_3$ remains zero until second order in perturbation theory.
Similarly, equation (14) is a third-order approximation because
the calculation of $S_4$ requires third-order theory.
GGRW compute values of $S_3$, $S_4$, and $S_5$ for a cold dark matter (CDM)
power spectrum smoothed with a Gaussian filter, and JBC derive
$S_3$ for power law spectra smoothed with Gaussian or top hat filters.
Bernardeau (in preparation)
derives $S_4$ for a power law spectrum and top hat filter.
Note that all of these values are for the smoothed {\it final} density fields,
so by combining them with the Edgeworth expansion one
incorporates the effects of gravitational evolution and smoothing
in the correct order.

Thus far we have considered the PDF of the mass density field, which can be
estimated from a galaxy redshift survey if one assumes that galaxies trace
mass, or if one incorporates an assumed model of ``biased'' galaxy formation
to describe the relation between the galaxy distribution and the mass
distribution (see further discussion in \S 5).
However, if the input data set is a peculiar velocity field,
it makes more sense to look at the velocity divergence,
$$
\theta \equiv \vec{\nabla}\cdot\vec{v}/H_0,
\eqno(\new)
$$
where $\vec{v}$ is the peculiar velocity and $H_0$ is the Hubble constant.
Our results above can be immediately generalized; one just adopts the
values of $S_{\ell}$ that are relevant for $\theta$ instead of $\delta.$
The computation of these moment ratios is discussed by
Bernardeau \etal (in preparation, hereafter BJDB), who give
values of $S_{3\theta}$
for power law spectra smoothed with Gaussian or top hat filters,
and by Bernardeau (in preparation), who gives $S_{4\theta}$ for
power law spectra smoothed with a top hat filter.

\bigskip

\noindent
{\bf 3. Measures of Asymmetry and Width \hfill}

In Gaussian models, second-order perturbation theory predicts
that $S_3\equiv\dcube/\dsqr^2$ should be a constant,
depending only on the shape of the power spectrum near the smoothing scale.
One can check whether the density fields derived from galaxy redshift
surveys and peculiar velocity surveys obey this relation in order to
test the hypothesis of Gaussian initial fluctuations (for studies along these
lines, see Bouchet \etal 1992b, 1993; Park 1991;
Silk \& Juszkiewicz 1991; Coles \& Frenk 1992).
A disadvantage of using the third moment $\dcube$
is that it is quite sensitive to the tails of the PDF, so it is subject to
sampling errors unless the survey volume is very large.  Recognizing
this problem, Nusser \& Dekel (1993) use the quantity $\dd$ instead of
$\dcube$ as a measure of asymmetry in the PDF.  In the Appendix,
we outline a direct perturbative calculation of $\dd$ for the unsmoothed
density field that evolves from smoothed initial conditions.
However, once one has computed the value of $S_3$ from perturbation theory,
one can derive the same result much more simply
using the Edgeworth approximation to the PDF.
Indeed, to calculate the first non-vanishing term in the asymptotic
expansion for $\dd$, it is sufficient to use the second-order
Edgeworth expansion (13):
$$
\dd = \sigma^2 \, \int_0^{\infty}\;\nu^2 \, [p(\nu) - p(-\nu)] \, d\nu
= {S_3\sigma^3\over 3} \, \int_0^{\infty}\; \nu^2H_3(\nu) \, \phi(\nu)
\, d\nu  + \; O(\sigma^5) \; ,
\eqno(\new)
$$
with $H_3(\nu) = \nu^3 - 3\nu$.
Equation (16) is accurate to $O(\sigma^5)$ because the additional
terms in the next order of the Edgeworth expansion are symmetric in $\nu$
and make no contribution to $\dd$.
Evaluating the simple integral on the right hand side of
equation (\eqref{1}), we find
$$
\dd = \sqrt{2\over 9\pi} \, S_3\sigma^3 \; + \; O(\sigma^5) \; .
\eqno(\new)
$$
This result is more general than the value for $\dd$ derived in
the Appendix ``from first principles'' because it is not
restricted to the unsmoothed field. The effects of a low pass
filter can be included simply by using the value of $S_3$ calculated from
perturbation theory for the smoothed final density contrast.
One can also incorporate the effects of shot noise, since the impact
of Poisson fluctuations on $S_3$ can be calculated easily.
Redshift-space distortions can be treated using the $S_3$
results of Bouchet \etal (in preparation), and biased galaxy
formation can be included using the results in \S 5 below
(see also Fry \& Gazta\~naga 1993).
Finally, from the above derivation it is clear how to compute the
moment $\tt$ of the smoothed velocity divergence; one
simply substitutes the appropriate value of $S_{3\theta}$ for $S_3$.

The quantity $\dd$ measures asymmetry of the PDF, in similar fashion
to the third moment $\dcube$.  A measure analogous to the
reduced fourth moment, $M_4=\dfour-3\sigma^4$, is the expectation value
of $|\delta|$, minus the contribution expected for a Gaussian PDF.
Nusser \& Dekel (1993) use this quantity, $\dmodr$, to
measure the width of the PDF relative to a Gaussian distribution.
We can compute its evolution in perturbation theory using the third-order
Edgeworth expansion (14):
$$ \eqalign{
  \dmod
   &= \sigma \, \int_0^{\infty}\,\nu \, [p(\nu)+p(-\nu)]\, d\nu \cr
   &= 2\sigma \, \left[\,
      \int_0^\infty \nu \phi(\nu) d\nu \, + \,
      {1\over 4!}S_4\sigma^2\int_0^\infty\nu H_4(\nu)\phi(\nu) d\nu \, + \,
      {10\over 6!}S_3^2\sigma^2\int_0^\infty\nu H_6(\nu)\phi(\nu) d\nu
      \, \right] \; .
}
\eqno(\new)
$$
Evaluating the integrals (the last two by parts, using equation [2])
and rearranging terms we find
$$
\dmodr = (2\pi)^{-1/2}(S_3^2-S_4) {\sigma^3 \over 12} \; + \; O(\sigma^5) \; .
\eqno(\new)
$$
Once again, we can compute this quantity for the smoothed final density
field or the smoothed velocity divergence by inserting the appropriate
values of $S_3$ and $S_4$ from perturbation theory.
The fourth moment weights the tails of the PDF heavily, but $\dmod$ responds
primarily to the width of the PDF near its peak, so a distribution with high
kurtosis (high $S_4$) tends to have a negative value of $\dmodr$.

For convenience in the sections that follow, we now introduce the notation
$$\eqalignno{
\tm3 &\equiv \dd, \qquad \qquad \qquad \ \ \;
\ts3  \equiv \sqrt{2\over 9\pi} \, S_3, &(\new) \cr
\tm4 &\equiv \dmodr, \qquad
\ts4  \equiv (2\pi)^{-1/2}(S_3^2-S_4)/12, &(\new) \cr
}
$$
in obvious analogy to the cumulants $M_3$, $M_4$ and the
normalized cumulants $S_3$, $S_4$.
In this notation, the predictions of perturbation theory combined with
the Edgeworth approximation are simply
$$\eqalignno{
\tm3 &= \ts3 \sigma^3 &(\new)\cr
\tm4 &= \ts4 \sigma^3. &(\new)\cr
}
$$

\bigskip

\noindent
{\bf 4. Comparison to \nbody\ Simulations}
\bigskip

\noindent
{\bf 4.1 Simulation Methods}

We can check the range of validity of our analytic approximations
by comparing their predictions to the results of fully non-linear,
cosmological \nbody\ simulations.  Here we examine simulations in which
the initial conditions are Gaussian with a power law power spectrum,
$P(k)=k^n$ with $n=-1$.  We have also analyzed simulations with
$n=0$; we will discuss these results briefly in \S 4.4.
We assume an Einstein-de Sitter
($\Omega=1$) background cosmology.

Our simulations use a particle mesh (PM) \nbody\ code written by
Changbom Park.  The code is described by Park
(1990; see also Park \& Gott 1991).  Its performance has been tested
against analytic solutions for non-linear pancake collapse and against
other PM codes, P$^3$M codes, and tree codes.  Since the perturbation
theory results described above must apply for sufficiently
low fluctuation amplitudes,
comparison of simulations to these results gives us an opportunity to
test the PM code in a new regime, that of weakly non-linear clustering.

We wish to examine behavior over a wide dynamic range, so we employ
large simulations that evolve $100^3$ particles on a $200^3$
force mesh.  The code uses a staggered-mesh technique (Melott 1986)
to achieve higher force resolution (by about a factor of two) than
a conventional PM code.
We advance the particle distributions to
expansion factor $a=1/128$ via the Zel'dovich approximation, then use
the PM code to integrate to $a=1$ in 127 timesteps of equal $\Delta a$.
We obtain output at expansion factors
$a=1/8,~1/4,~1/2$ and 1.
To analyze the final density fields, we first bin
the particles onto a $100^3$ grid using a cloud-in-cell (CIC) weighting scheme,
then smooth this field by Fourier convolution with Gaussian filters of varying
smoothing lengths $r_s$.  We normalize the initial power spectrum so
that at the final output time ($a=1$)
the r.m.s. fluctuation predicted by linear theory on a Gaussian smoothing scale
of $r_s=2$ cells is $\sigma=2$.
At each output time we analyze the density
fields with smoothing lengths $r_s=2$, 4, and 8 cells.
Each of these cells is twice the linear size of the cells used
for force calculations in the simulations, so the
effective gravitational softening length is quite a bit smaller than
even our smallest smoothing scale.
We have run eight simulations with independent initial conditions.

The great advantage of adopting initial conditions
with power law power spectra is that one can check the reliability of the
numerical results by looking for self-similar behavior (see discussion by
Efstathiou \etal 1988).
Neither the form of the initial spectrum nor the $\Omega=1$ cosmology
introduces a preferred scale, so at any time only the amplitude of fluctuations
is available to define a characteristic radius.  Statistical properties
of the density field smoothed at a particular scale should depend only
on the r.m.s. fluctuation amplitude on that scale, regardless of whether
structure
on that scale is linear, weakly non-linear, or strongly non-linear.
For $\Omega=1$ and $n=-1$, the r.m.s.
linear fluctuation amplitude is proportional to the expansion factor and
inversely proportional to the smoothing length, so with our normalization
$\sigma(a,r_s)=4a/r_s$.  (If we adopt the standard procedure of
matching the simulations' mass fluctuations to observed galaxy count
fluctuations, then the implied physical size of the simulation cube is
$\sim 60a^{-1}~h^{-1}$ Mpc on a side.)
To the extent that the simulations are correct,
we expect statistical results to depend on $a$ and $r_s$ only through
the ratio $a/r_s$, and our choices of output times and smoothing lengths
provide a number of ``degenerate'' combinations with which to test for
this self-similar scaling.  Numerical parameters like the force resolution,
particle density, and box size remain fixed in simulation units, so numerical
artifacts that reflect the simulation's finite dynamic range should violate
this scaling.  For the measures that we examine in this paper, our simulations
obey the self-similar scaling extremely well, except for some modest
finite-volume effects, which are noticeable at the largest smoothing length,
$r_s=8$.

Two features of our simulations are essential to obtaining the excellent
agreement between \nbody\ and perturbative results illustrated below.
The first is a high initial redshift, so that the physical volume
represented by the simulation grows by a substantial factor before the
first output time.  If the expansion factor is too small, then the
\nbody\ results will contain transients that reflect the use of the
Zel'dovich approximation to generate initial conditions.  The Zel'dovich
approximation yields incorrect values for the moments of the density
field, primarily because it conserves momentum only to first order.
For skewness this failure is relatively modest; with a top hat filter
and $P(k) \propto k^n$ power spectrum, for instance, the Zel'dovich
approximation yields $S_3=4-(n+3)$ instead of the value
$S_3=34/7 - (n+3)$ obtained from rigorous, second-order perturbation
theory (JBC).  However, the approximation deteriorates catastrophically
for higher order cumulants (Grinstein \& Wise 1987; JBC).
Furthermore, it fails badly even at the skewness level when applied
to the divergence of the velocity field; with a top hat filter and
power law spectrum, the Zel'dovich approximation predicts
$S_{3\theta} \equiv \tcube/\tsqr^2 = \Omega^{0.6}(n+1)$, while
perturbation theory yields $S_{3\theta}=\Omega^{0.6}(n-5/7)$,
i.e.\ a skewness of the opposite {\it sign} for $-1 < n < 5/7$
(BJDB).
We therefore believe that
calculations of the PDF based on the Zel'dovich approximation
should be treated with caution, though they may yield a useful
qualitative picture in some cases.

The second essential feature of these simulations, for our application,
is the rather fine ($200^3$) force mesh used during
dynamical evolution.  We first carried out these experiments using
a $100^3$ force mesh, and we found $\sim 20\%$ discrepancies between the
\nbody\ and analytic results for skewness of the density field at
low variance, along with comparable violations of self-similar scaling
in the simulations themselves.  The high-resolution
mesh is needed precisely because we are investigating {\it weakly}
non-linear clustering.  PM codes make significant errors in the linear
evolution of Fourier modes with wavelengths of a few mesh cells
(Bouchet, Adam \& Pellat 1985).  However, once clustering becomes
fully non-linear, structure on the mesh scale is determined mainly
by the collapse of modes that initially had large wavelengths, and
which were therefore evolved accurately through the linear
regime (Little, Weinberg \& Park 1991; Moutarde \etal 1991).
We thus expect reliable results from a PM code fairly close to the
mesh scale provided that the r.m.s.\ fluctuation exceeds unity on
the scale of a few mesh cells.  Most \nbody\ studies operate
in this regime, and we believe that it is this transfer of power
from large scales to small scales that accounts for the
relatively good agreement between different types of cosmological
\nbody\ codes found by Weinberg \etal (unpublished comparison).
Our present investigation of the weakly non-linear regime requires
a high-resolution mesh because we are interested in the regime
of small fluctuation amplitudes.  Note that it is specifically
the mesh used to compute forces during dynamical
evolution that is relevant to these considerations, and that a
staggered-mesh PM scheme offers significant advantages over a
traditional PM scheme (Melott 1986; Melott, Weinberg \& Gott 1988;
Park 1990).

\topinsert
\capskip
\centerline{
\epsfxsize=3.5truein
\epsfbox[116 306 464 728]{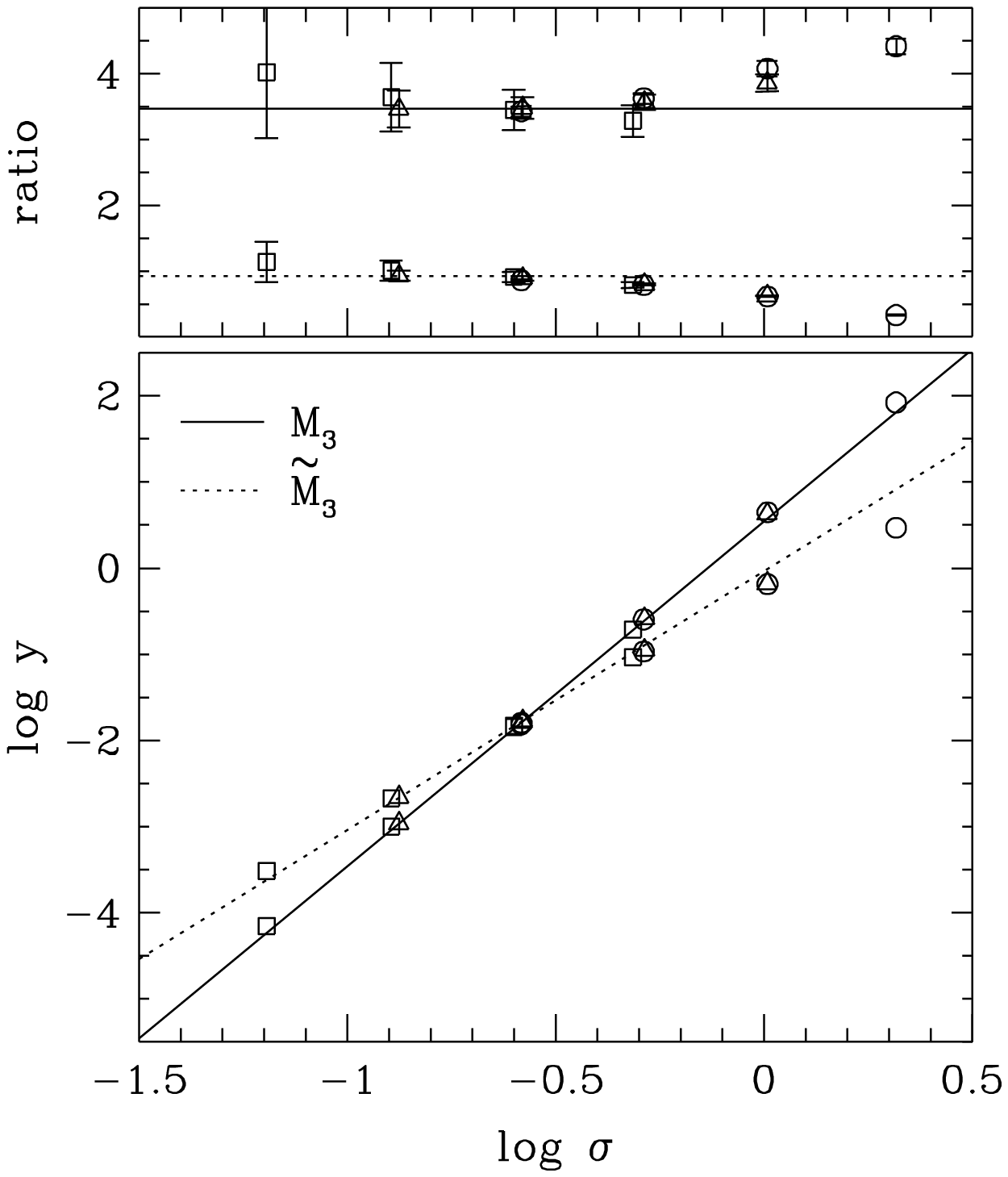}
}
\medskip
{\eightpoint
\noindent {\bf Figure 2} --- Evolution of the asymmetry measures
$M_3\equiv\dcube$ and $\tm3\equiv\dd$.
Logarithms are base-10 in this and all subsequent figures.
In the lower panel, the solid line shows the
prediction of second-order perturbation theory, $M_3 = S_3 \sigma^4$,
using the value $S_3=3.47$
appropriate to an $n=-1$ power spectrum and Gaussian smoothing filter.
The dotted line shows the relation computed from the second-order
Edgeworth approximation, $\tm3 = \ts3\sigma^3$.
Points show measurements from the density fields of the \nbody\ simulations,
with smoothing lengths of 2, 4, and 8 cells (circles, triangles, and squares,
respectively).  We use the same set of symbols for $M_3$ and $\tm3$,
but they can be distinguished by their close fits to the corresponding
analytic predictions.
For closer inspection, the upper panel plots the ratios $M_3/\sigma^4$
(top points) and $\tm3/\sigma^3$ (bottom points), with horizontal lines
representing the analytic predictions.  Error bars mark the $1\sigma$
numerical uncertainty, i.e.\ the run-to-run dispersion of the eight
independent simulations divided by $7^{1/2}$.
Perturbative and \nbody\ results agree to within this uncertainty
when $\sigma$ is small, as expected. \nbody\ results for $M_3$ remain
remarkably close to the perturbation theory prediction even when $\sigma=2.$
There are no free parameters to either of these ``fits.''
}
\smallskip
\textskip
\endinsert

\bigskip
\noindent
{\bf 4.2 Moments and PDF of the Density Field}

Figure 2 compares moments of the $N$-body density fields to the predictions
of second-order perturbation theory.
The solid line in the lower panel shows the perturbative relation
$M_3 \equiv \dcube=S_3\sigma^4$, where we have used the value
$S_3=3.47$ appropriate for an $n=-1$ power spectrum
and Gaussian smoothing filter (JBC).  The dotted
line shows the relation
$\tm3 \equiv \dd = \ts3 \sigma^3$, derived in \S 3.
Points show corresponding results from the \nbody\ density
fields (averaged over the 8 runs), with Gaussian smoothing lengths of
2 cells (circles), 4 cells (triangles), and 8 cells (squares).
We use the same set of symbols for both $M_3$ and $\tm3$,
but the two quantities can be distinguished because of their close fits to the
corresponding analytic results.  Concentrating first on the \nbody\ results,
we see that the circles and triangles agree almost perfectly within their
range of overlap, indicating that at these smoothing lengths the
simulations obey the expected self-similar scaling.  Values for $r_s=8$
(squares) agree quite well, but all three moments have
slightly lower amplitudes, because the smoothing length is large enough that
the absence of power on scales larger than the fundamental mode of the
simulation cube has become a significant effect.

The agreement between the \nbody\ results and the second-order prediction
for $M_3$, in a plot spanning six orders of magnitude, is rather
spectacular.
For closer inspection, the top set of points in the upper
panel of Figure 2 presents the ratio $M_3/\sigma^4$ as a function of
$\sigma$, and the horizontal solid line shows the perturbation theory
prediction, $M_3/\sigma^4 = 3.47$.
Error bars mark the $1\sigma$ uncertainty in the mean from the eight $N$-body
simulations, i.e.\ the run-to-run dispersion in the value of this ratio
divided by $7^{1/2}$.  These error bars represent the uncertainty of our
numerical estimates, and they increase with smoothing length
because the number of independent smoothing volumes per simulation decreases.
The analytic and numerical results agree to within the
statistical error of the \nbody\ simulations for $\sigma\leq 1/2$.
For higher $\sigma$ the $N$-body results climb above the perturbation
theory prediction, but they still
agree to within 15\% at $\sigma=1$ and 25\% at $\sigma=2$, well into the
regime where second-order perturbation theory should break down.
Similar behavior has been seen in
observational data (Bouchet \etal 1993 and references therein)
and in other numerical simulations
(Lucchin \etal 1993 and references therein).
In many cases
the near proportionality between $M_3$ and $\sigma^4$ extends still
further, into the strongly non-linear regime.
On the observational side,
proportionality of moments is implied by the
well-known result that the reduced 3-point correlation
function at small separations
can be expressed accurately as a sum of pairwise products of
the two-point correlation function,
$\zeta_{123}=Q(\xi_1\xi_2 + \xi_2\xi_3 + \xi_3\xi_1)$ (Peebles 1980).
Among the \nbody\ results, including our own,
there is a rough
consensus that the ratio $M_3/\sigma^4$
rises somewhat above the perturbation theory value when $\sigma$
exceeds unity, but the change is a modest one.
While perturbation theory shows that $M_3$ should scale with
$\sigma^4$ when clustering is weak, there is as yet no fundamental
explanation for the continuation of this relation into the strongly
non-linear regime.

Figure 2 also shows agreement between the \nbody\ results and
the Edgeworth calculation of $\tm3$ for small $\sigma$.
However, in this case the analytic predictions begin to separate from the
numerical results more quickly.  The perturbative calculation overestimates
the \nbody\ values of $\tm3$ by 15\% at $\sigma=1/2$,
30\% at $\sigma=1$, and 60\% at $\sigma=2$.
Even these errors are
not so large when one recalls that the predicted value of $\tm3$
is growing by a factor of 8 for each factor of 2 increase in $\sigma$.
Positive and negative fluctuations grow at different rates in the
non-linear regime, and since the $M_3$ and $\tm3$ moments
weight extreme values differently, it would be virtually impossible for
perturbative calculations of {\it both} quantities to remain accurate
once $\sigma$ approaches or exceeds one.

\topinsert
\capskip
\centerline{
\epsfxsize=3.5truein
\epsfbox[116 306 464 728]{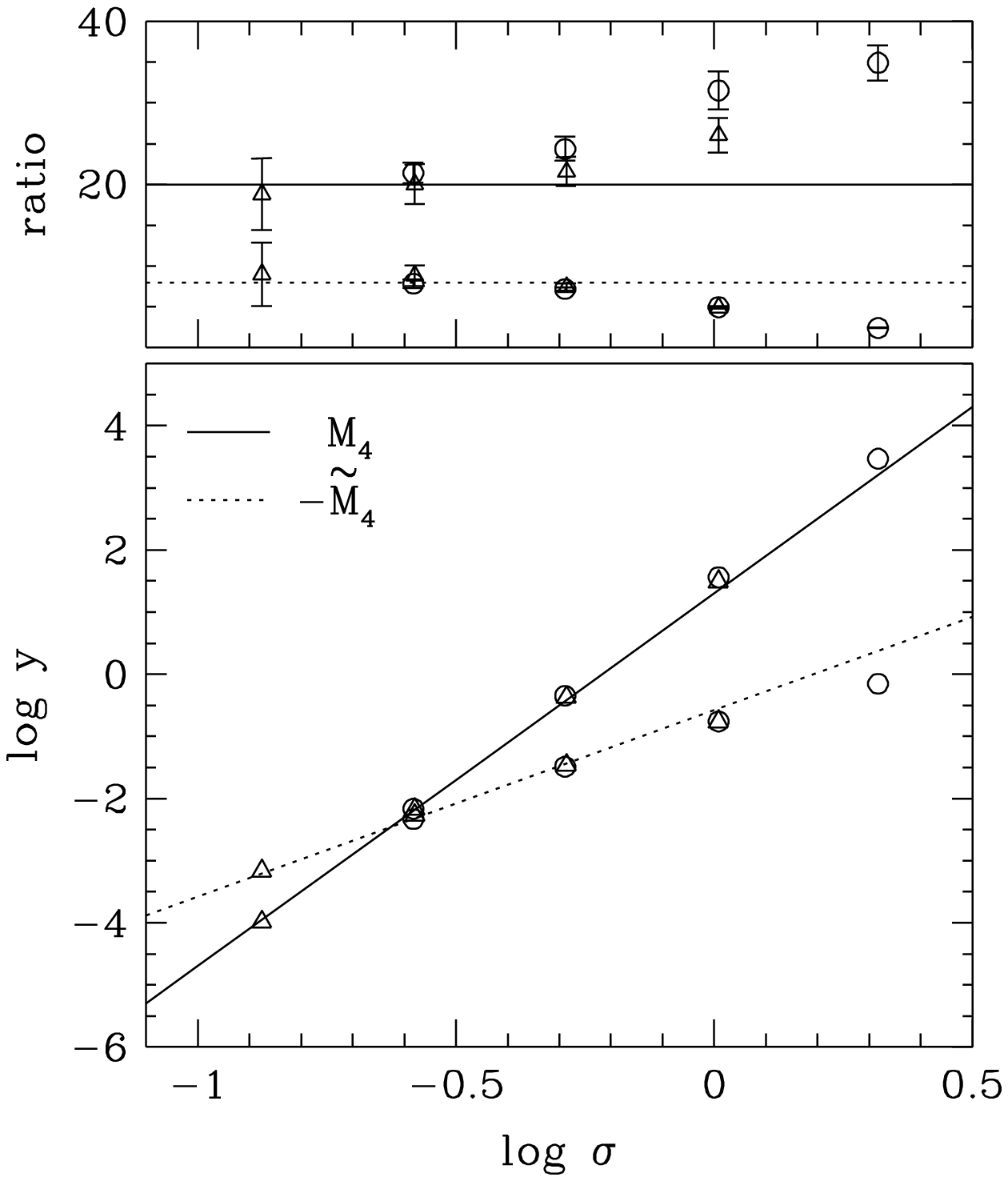}
}
\medskip
{\eightpoint
\noindent {\bf Figure 3} ---
Evolution of the width measures
$M_4 \equiv \dfour-3\sigma^4$ and $-\tm4 \equiv (2/\pi)^{1/2}\sigma-\dmod$.
In the lower panel,
the solid line shows the prediction of third-order perturbation theory,
$M_3 = S_4\sigma^6$.
The dotted line shows the relation computed from the third-order Edgeworth
approximation, $-\tm4 = -\ts4\sigma^3$ (note that $\ts4$ is negative).
Points show measurements from the density fields
of the \nbody\ simulations, with smoothing lengths of 2 cells (circles)
and 4 cells (triangles).
The upper panel displays the ratios $M_4/\sigma^6$ and
$30 \times (-\tm4/\sigma^3)$, with the factor 30 chosen to make the
results easily visible on this plot.
We do not have an analytically calculated value
of $S_4$, so we have chosen a value $S_4=20$ that provides a good eye-fit
to the \nbody\ results for $M_4$ at low $\sigma$.
This constant is the only
free parameter in these ``fits''; the slopes of the power laws
are determined by perturbation theory, and the value of $\ts4$ is fixed
by the choice of $S_4$.
}
\smallskip
\textskip
\endinsert

In Figure 3 we plot the width measures $M_4$ and $\tm4$ against $\sigma$.
The solid line in the lower panel
shows the prediction of third-order perturbation theory
for the fourth-order cumulant:
$M_4 \equiv \dfour-3\sigma^4 = S_4\sigma^6.$
Points show the \nbody\ results for
$r_s=2$ (circles) and $r_s=4$ (triangles).
The fourth moment is very noisy at $r_s=8$, and we
do not show results for this smoothing length.
In the upper panel, the top set of points shows the ratio of $M_4$ to
$\sigma^6$, with error bars representing the $1\sigma$ uncertainty from
the simulations and the horizontal solid line representing the prediction
$M_4/\sigma^6=\,$constant, from perturbation theory.
We do not have an analytically
calculated value of $S_4$ for this spectrum and smoothing filter, so
we have treated it as a free parameter and chosen a value $S_4=20$ that
provides a good eye-fit to the low-$\sigma$ points in Figure 3.
This value is in reasonable accord with GGRW's computation
for a CDM power spectrum; they find $S_4 \approx 20$ on a smoothing
scale where the slope of the CDM power spectrum is $n \sim -1$.

The dotted line in the lower panel of Figure 3 displays the relation
$-\tm4\equiv (2/\pi)^{1/2}\sigma-\langle|\delta|\rangle=-\ts4\sigma^3$
predicted by the Edgeworth approximation
(equation 23; note that $\ts4$ is negative).
We compute the value of $\ts4$ using
$S_4=20$, determined from the fit to the $M_4$ vs.\ $\sigma^6$
plot, and $S_3=3.47$, determined analytically, so there is no
freedom to adjust the height or slope of the dotted line.
Points show the corresponding $N$-body results.
In the upper panel, the bottom set of points and the dotted line show the
ratio $-\tm4/\sigma^3$ for the simulations and the Edgeworth
approximation, respectively.
We have multiplied all the values
and the error bars by a factor of 30 in order to make them clearly visible
on this plot.  The results of Figure 3 are similar to those of Figure 2.
For both $M_4$ and $\tm4$,
the \nbody\ points follow the power laws predicted by
perturbation theory when $\sigma$ is small.
The cumulant $M_4$ remains close to the perturbation theory prediction
even for $\sigma \sim 1-2$.  The Edgeworth approximation for
$\tm4$ stays fairly accurate up to $\sigma=1$, but it overestimates
the numerical results significantly at $\sigma=2$.

\topinsert
\capskip
\centerline{
\epsfxsize=5.5truein
\epsfbox[45 198 576 738]{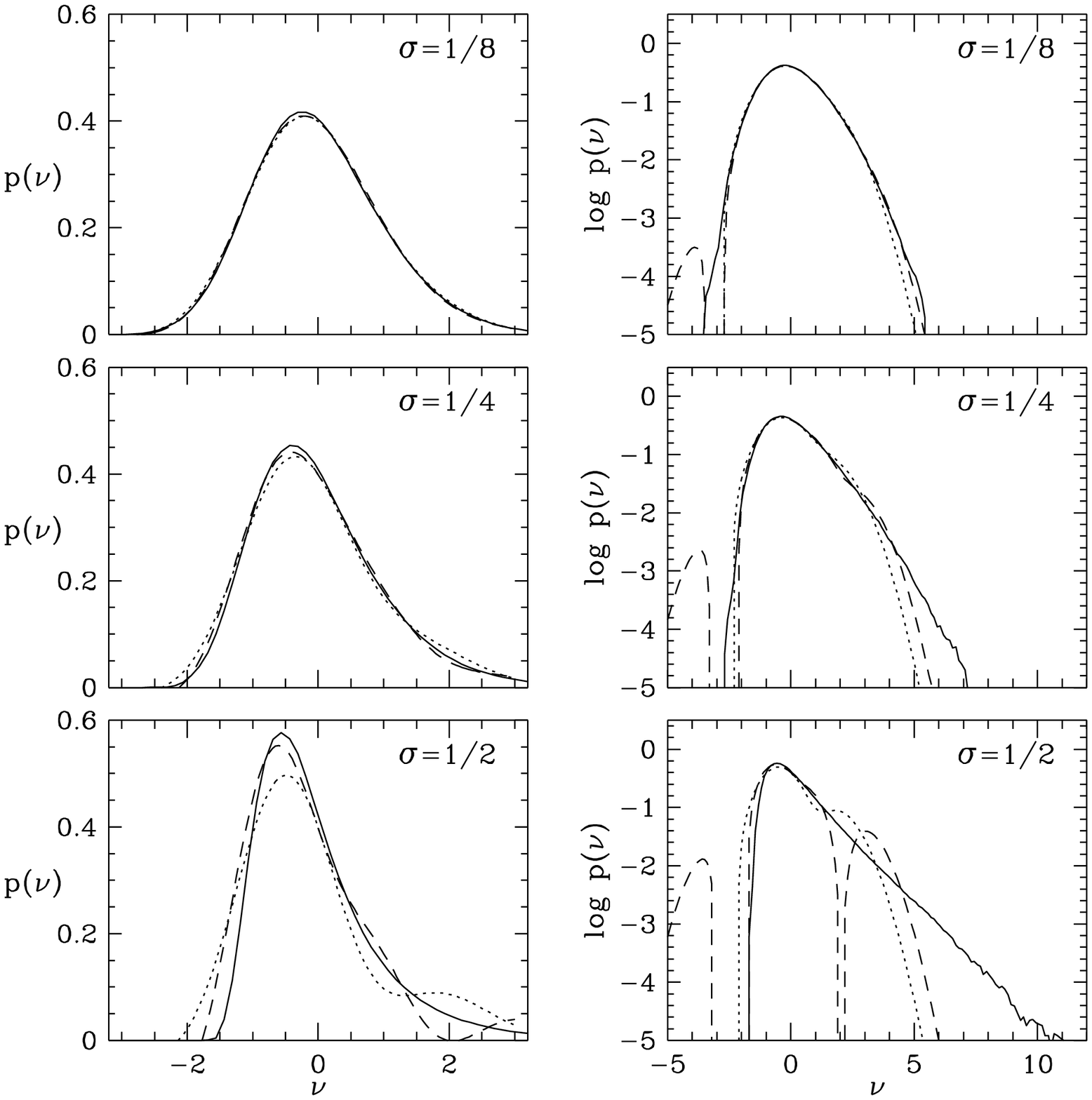}
}
\medskip
{\eightpoint
\noindent {\bf Figure 4} ---
Probability distribution functions (PDFs) evolving from
Gaussian initial conditions.  Left hand panels are linear plots, emphasizing
behavior near the peak of the PDF.  Right hand panels are logarithmic,
emphasizing the tails.  Solid lines show PDFs measured from the
smoothed density fields of the \nbody\ simulations when the r.m.s.\ fluctuation
is $1/8$ (top), $1/4$ (middle), and $1/2$ (bottom).  Dotted lines show the
second-order Edgeworth approximation to the PDF, equation (13).
Dashed lines show the third-order Edgeworth approximation, equation (14).
Both approximations work well for $\sigma=1/4$ and begin to break down
when $\sigma=1/2$.
}
\smallskip
\textskip
\endinsert

Figure 4 brings us to the central issue of this paper, the
overall shape of the PDF.
The solid lines show PDFs of the \nbody\ density fields
for three different values of the r.m.s. fluctuation amplitude, $\sigma=1/8$
(top panels), $\sigma=1/4$ (middle panels), and $\sigma=1/2$ (bottom panels).
Left hand panels plot $p(\nu)$ against $\nu\equiv\delta/\sigma$ and
emphasize behavior near the peaks of the distributions.  Right hand panels
plot ${\hbox{log}}_{10}p$ against $\nu$ in order to display the tails of the
distributions.  We average the results from the eight \nbody\ density fields
analyzed with smoothing length $r_s=4$; the curves for different values
of $\sigma$ are obtained from output expansion factors $a=\sigma$.
We obtain identical results from other combinations of $a$ and $r_s$
that have the same $\sigma$, except for minor finite-volume effects on the
extreme tails of the distributions.

The dotted lines in Figure 4 show the second-order Edgeworth approximation
to the PDF (equation 13), i.e.\ including only the Gaussian and
skewness terms of the Edgeworth series.  The approximation does
very well at $\sigma=1/8$, and it remains accurate at $\sigma=1/4$ except
that the positive tail falls too rapidly for $\nu > 4$.  At $\sigma=1/2$
the approximation is breaking down, though it still oscillates around the
true PDF in a reasonable way.
The dashed lines show the effect of including the third-order
term of the Edgeworth expansion (equation 14).
We use the value $S_4=20$ estimated from the moment ratios in Figure 3.
This third-order approximation is somewhat more accurate than the
second-order approximation at each value of $\sigma$.
This is the behavior that we expect, since our approximation is a power
series expansion in $\sigma$.
The third-order term provides higher accuracy, and it slightly extends
the useful range of the Edgeworth expansion.
However, the third-order approximation begins to break down fairly soon
after the second-order approximation fails; the Edgeworth series is a
powerful approach for describing the first deviations from Gaussian
fluctuations, but it cannot take one into the deeply non-linear regime.
Also, as Figure 4 illustrates, the Edgeworth approximation to the PDF
is not positive-definite, though it does integrate to unity at any order,
and it is positive-definite when $\sigma \ll 1$.
We have not investigated the effects of including higher-order terms
of the Edgeworth series.
If one uses the Gram-Charlier series (1) instead of the Edgeworth series (12),
then adding the third-order term {\it degrades} the fit to the \nbody\
PDF.  This failure is not surprising, since the Gram-Charlier series
is not a proper asymptotic expansion.

\topinsert
\capskip
\centerline{
\epsfxsize=3.5truein
\epsfbox[116 306 464 728]{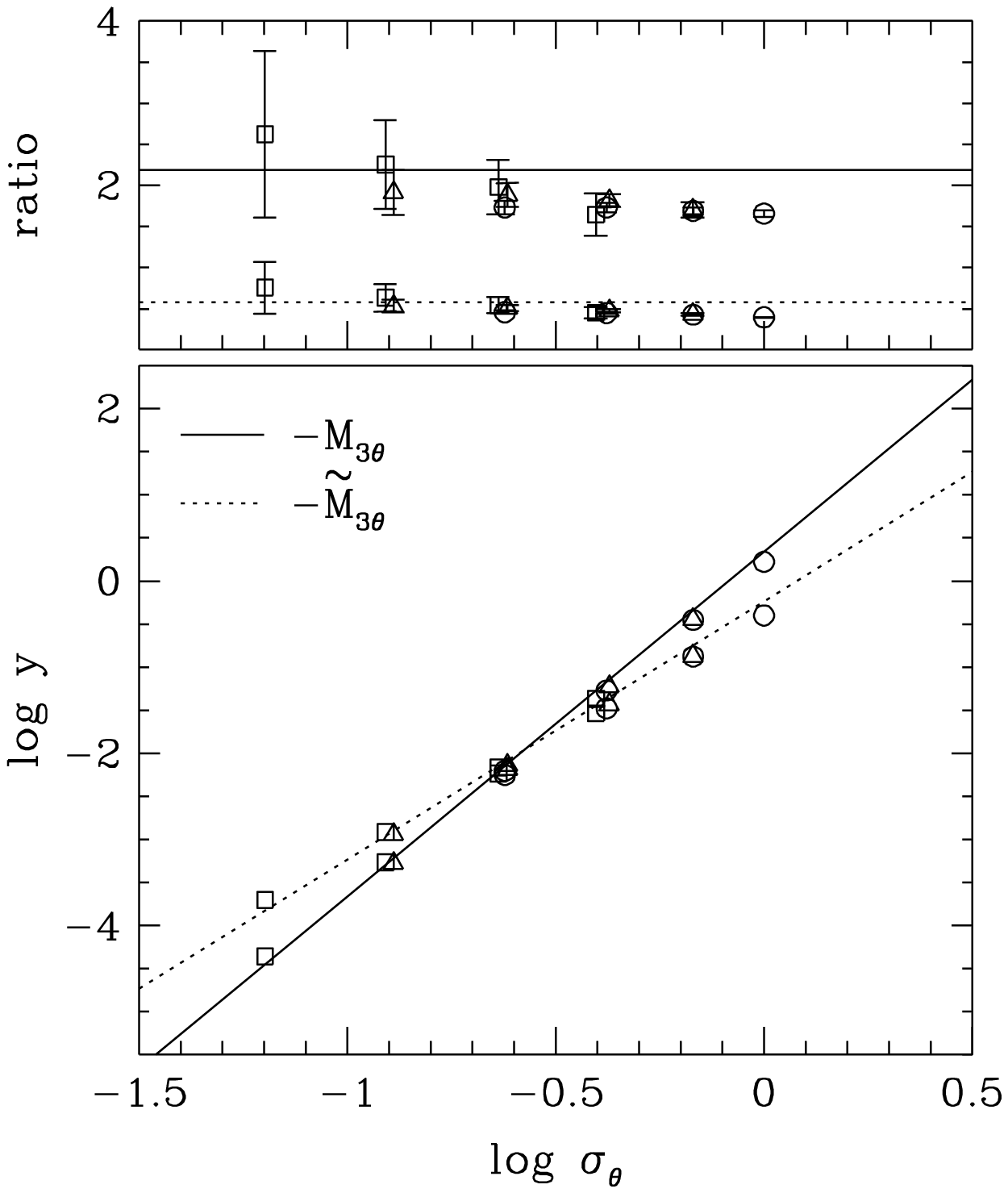}
}
\medskip
{\eightpoint
\noindent {\bf Figure 5} ---
Evolution of $-M_{3\theta}\equiv-\tcube$ and
$-\tm{3\theta}\equiv -\langle \theta |\theta| \rangle $, where
$\theta \equiv \vec{\nabla}\cdot\vec{v}/H_0$ is the divergence of the
peculiar velocity field.  The format is the same as Figure 2.  We use the
value $S_{3\theta}=-2.19$ appropriate to an $n=-1$ power spectrum and Gaussian
smoothing filter, so there are no free parameters to either of these ``fits.''
}
\smallskip
\textskip
\endinsert

\bigskip
\noindent
{\bf 4.3 Moments and PDF of the Velocity Divergence}

We now turn our attention to the velocity field, or, more specifically,
to its divergence.  Perturbative calculations describe a smoothed velocity
field in which each volume element has equal weight.  However, \nbody\
simulations have a finite number of particles, not a continuum,
and the simulation velocity field is known only at the particle locations.
When the clustered particle distribution is binned onto a grid, some cells
may be empty, but this does not mean that the velocities in those cells are
zero, just that they are undetermined.
It is straightforward to define a mass-weighted,
smoothed velocity field as the ratio of the smoothed momentum field to the
smoothed density field,
but the results may not agree with perturbative
calculations because of the difference in definitions.
[A velocity field defined on a grid by averaging the velocities of the
particles in each cell is, in fact, a mass-weighted velocity field with
a cubic cell smoothing kernel.]
To get around this
problem, we define the velocity field smoothed on scale $r_s$ by first
computing a mass-weighted velocity field on a grid with a Gaussian
smoothing length $r_s/2$, then smoothing this field in a volume-weighted way
with a smoothing length $r_s^{\prime} = (r_s^2-r_s^2/4)^{1/2}$.
The first step assigns sensible, non-zero
velocity values to all cells, but since smoothing lengths add in quadrature,
the second, volume-weighted smoothing dominates the final result,
and the combined smoothing length is equal to $r_s$.
When density fluctuations are small, this procedure is equivalent to
simple, volume-weighted smoothing of the velocity field.
Once the smoothed velocities have been calculated, we compute the PDF and
moments of the divergence field
$\theta \equiv \vec{\nabla} \cdot \vec{v}/H_0.$
In the linear regime, $-\theta$ is equal to the density contrast $\delta$, but
the non-linear evolution of the two fields is different even at second order.

Figure 5 plots $-M_{3\theta}\equiv-\tcube$ and $-\tm{3\theta}\equiv-\tt$
against $\sigma_\theta\equiv\tsqr^{1/2}$, in the same format as Figure 2.
The PDF of the velocity divergence has negative skewness,
because in the non-linear regime the inflows to high density regions
are faster than the outflows from low density regions.
For the analytic
predictions we use the coefficient $S_{3\theta}=-2.19$ appropriate to the
velocity divergence field (BJDB).
Once again, the \nbody\ results obey
the expected self-similar scaling except for minor finite-volume
discrepancies at $r_s=8$.  The horizontal spacing between points decreases
steadily, indicating that $\sigma_\theta$ is growing more
slowly than predicted by linear theory.  By the final time, the r.m.s.
fluctuations at 2 and 4 cells are $\sigma_\theta(2)=1.0$ and
$\sigma_\theta(4)=0.68$,
compared to the linear theory values of 2.0 and 1.0.
The r.m.s. fluctuation of the {\it density}
field, on the other hand, grows at almost exactly the linear rate
(see Figure 2).

\topinsert
\capskip
\centerline{
\epsfxsize=3.5truein
\epsfbox[116 306 464 728]{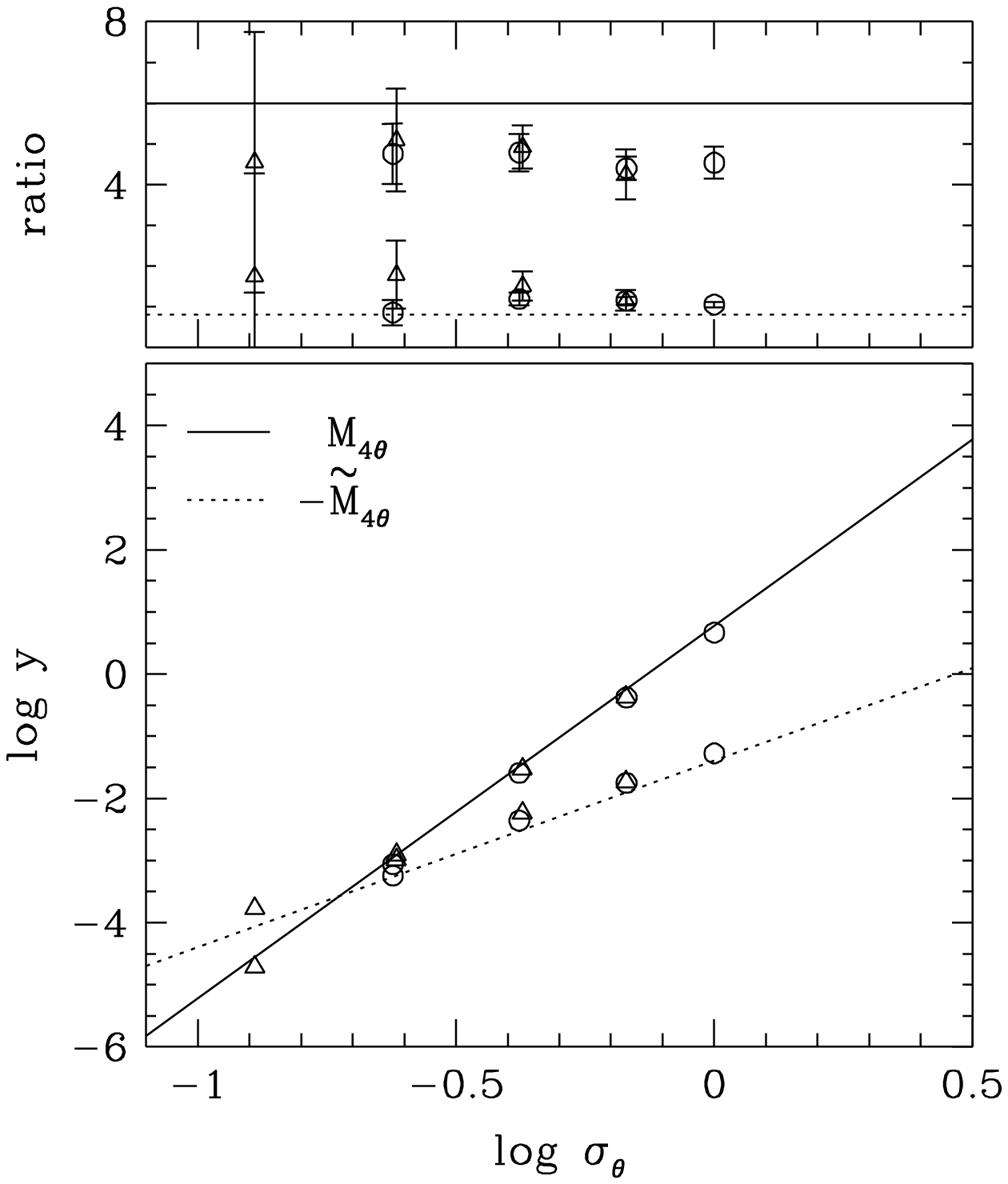}
}
\medskip
{\eightpoint
\noindent {\bf Figure 6} ---
Evolution of $M_{4\theta}$ and $-\tm{4\theta}$, in the same
format as Figure 3.  In the upper panel, we multiply the ratios
$-\tm{4\theta}/\sigma_\theta^3$ (bottom set of points and
horizontal dotted line)
by a factor of 20, to make the results easily visible on this plot.
We do not have an analytically calculated value of $S_{4\theta}$,
so we have chosen a value $S_{4\theta}=6.0$ that provides a
reasonable eye-fit to the \nbody\ results.  However, the $M_{4\theta}$ results
alone suggest a somewhat lower value ($S_4 \approx 4.5$), while the
$\tm{4\theta}$ results alone suggest a somewhat higher value
($S_4 \approx 6.5$).  The slopes of the power laws are determined by
perturbation theory, and the relation between $S_{4\theta}$ and
$\ts{4\theta}$ is determined by equation (21), so the value of $S_{4\theta}$
is the only free parameter in these ``fits.''
}
\smallskip
\textskip
\endinsert

\topinsert
\capskip
\centerline{
\epsfxsize=5.5truein
\epsfbox[45 198 576 738]{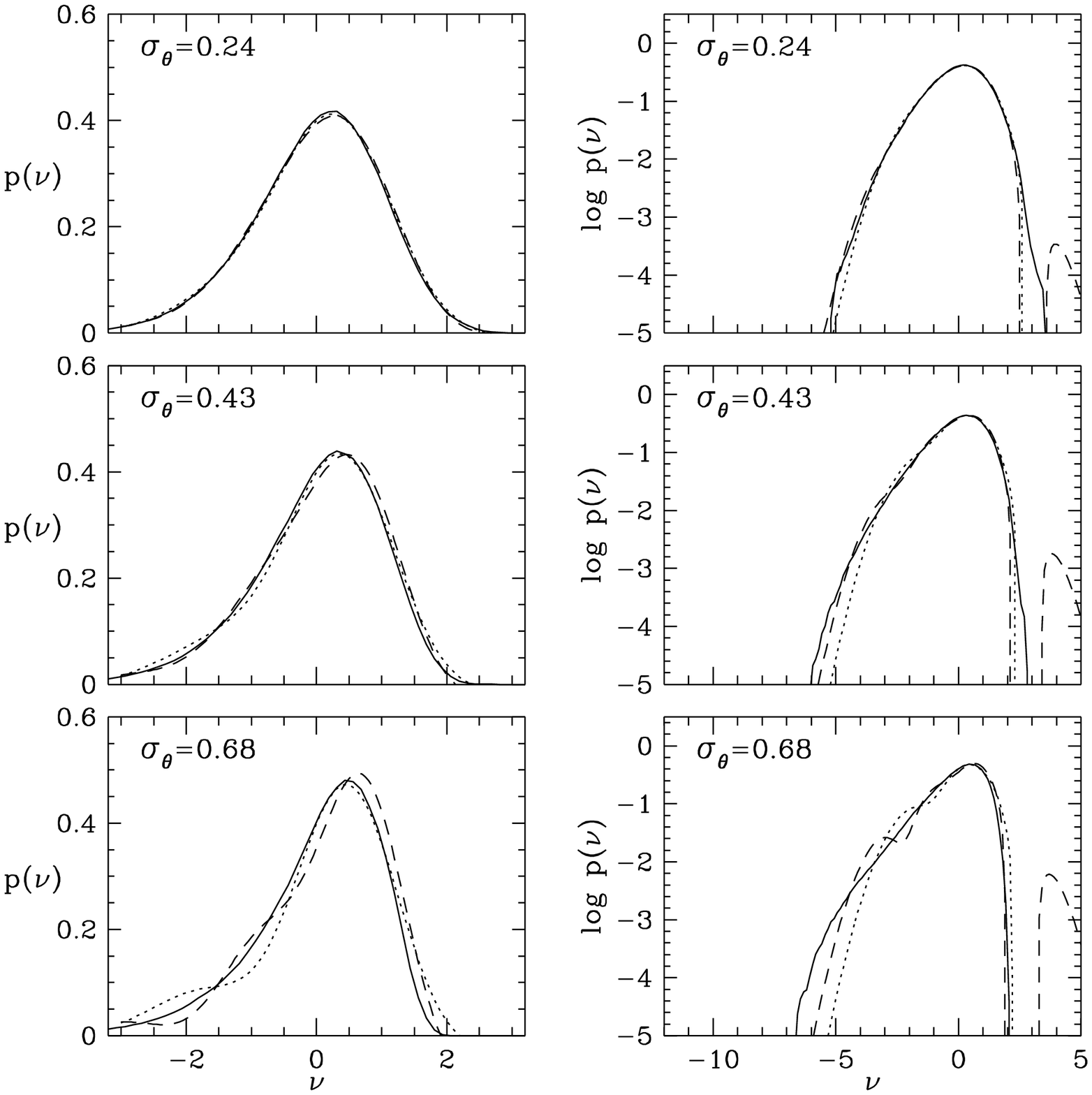}
}
\medskip
{\eightpoint
\noindent {\bf Figure 7} ---
Evolving PDFs of the velocity divergence, in the same format as
Figure 4.  Solid lines show PDFs measured from smoothed velocity fields of
the \nbody\ simulations, when the r.m.s.\
fluctuation of the velocity divergence
is 0.24 (top), 0.43 (middle), and 0.68 (bottom).  Dotted and dashed lines
show the second- and third-order Edgeworth approximations, respectively.
}
\smallskip
\textskip
\endinsert

Figure 5 demonstrates agreement between the \nbody\ and perturbative
calculations for low $\sigma_\theta$.
With the $n=-1$ power spectrum and Gaussian filter used here, the
Zel'dovich approximation predicts a skewness that is lower than
these values by nearly a factor of 8 (BJDB).
The second-order predictions remain close to the \nbody\ results
even as $\sigma_\theta$ approaches 1,
overestimating the numerical values of $-M_{3\theta}$ and $-\tm{3\theta}$
by about 15\%, 20\%, and 25-30\% for $\sigma_\theta=0.42$, 0.68, and 1.0
respectively (corresponding to linear theory values of $\sigma_\theta=0.5,$
1.0, and 2.0).
Figure 6 shows corresponding results for
$M_{4\theta}\equiv\tfour-3\sigma_\theta^4$ and
$-\tm{4\theta}\equiv (2/\pi)^{1/2}\sigma_\theta-\langle|\theta|\rangle$,
in the same format as Figure 3.
In the upper panel, we multiply the ratios $-\tm{4\theta}/\sigma_\theta^3$
by a factor of 20 to make them clearly visible on this plot.
We do not have an analytically calculated value of $S_{4\theta}$, and
we have chosen the value $S_{4\theta}=6$ on the basis of an eye-fit to
the \nbody\ points in Figure 6.  Small changes in the value of $S_{4\theta}$
induce large logarithmic changes in the value of
$\ts{4\theta}\propto S_{3\theta}^2-S_{4\theta}$ (equation 21),
because it is in a critical
region near zero.  If we were fitting the $\tm{4\theta}/\sigma_\theta^3$
ratio alone, we would probably
choose $S_{4\theta} \approx 6.5$, but the results for
$M_{4\theta}$ seem to indicate a lower value, $S_{4\theta} \approx 4.5$.
We do not know the cause of this mild discrepancy, but we would guess that
it reflects inaccuracies and statistical uncertainties
in the numerical results, the significance of
which is amplified by being in the critical region
$S_{4\theta}\approx S_{3\theta}^2$.
An analytically computed value of $S_{4\theta}$ might clarify
this issue, but the required integrations are rather forbidding.

Figure 7 compares the Edgeworth approximation for the PDF of $\theta$
to the PDFs measured from the \nbody\ simulations, in a format similar to
Figure 4.  Solid lines show the \nbody\ PDFs with a smoothing length of
4 cells at $a=1/4$, $1/2$, and 1 (top to bottom).  The values of the
r.m.s.\ fluctuation $\sigma_\theta$ are listed in each panel.
Dotted lines shows the second-order Edgeworth
approximation, equation (13) with $S_{3\theta}=-2.19$.
Dashed lines show the third-order approximation, equation (14)
with $S_{4\theta}=6$.
The PDF of the
velocity divergence develops a non-Gaussian shape more slowly than the PDF
of the density field, and the second-order Edgeworth approximation
remains accurate further into the non-linear regime, as one can see
by comparing to Figure 4 (where the r.m.s. fluctuations are actually lower).
It comes as no great surprise that the accuracy of the second-order Edgeworth
expression is closely related to the magnitude of the skewness,
$|S_3\sigma|$, or that the approximation begins to break down when
$|S_3\sigma|$ exceeds one.
The third-order term of the expansion generally improves the behavior
of the negative $\theta$ tail, but overall this term appears to be less
useful for the velocity divergence than it is for the density field
(compare to Figure 4).

\bigskip
\noindent
{\bf 4.4 Spectral Dependence of the PDF}

We have carried out all of the above comparisons for similar simulations
with a white noise ($n=0$) initial power spectrum.  We do not show the
results here, but the agreement between the perturbative and \nbody\
calculations is at least as good, and in some cases even better,
{\it provided} that one uses the values $S_3=3.14$ and $S_{3\theta}=-1.67$
appropriate to an $n=0$ spectrum (JBC; BJDB).  This leads us to an interesting
theoretical point.  Kofman \etal (1993) compute the density PDF that evolves
from Gaussian initial conditions using the Zel'dovich approximation.
In their calculation the shape of the PDF depends only on the r.m.s.\
fluctuation amplitude, $\sigma$, and it is independent of the shape of the
power spectrum.
Strictly speaking, Kofman {\it et al.}'s technique
describes the PDF of an unsmoothed density field evolving
from smoothed initial conditions, and if we used the second-order
Edgeworth approximation for this PDF we would also find a spectrum-independent
result, since the coefficient $S_3=34/7$ is independent of the power
spectrum in the absence of smoothing.
For the case of a smoothed {\it final} field,
the second-order Edgeworth PDF depends on both the r.m.s.\ fluctuation and
the shape of the power spectrum, but only through the combination
$S_3\sigma$.  This single parameter
tells us how to relate the predicted
PDFs for different power spectra, and how to
relate the PDF of the density field to that of the velocity divergence.

We can conjecture that the shape of the PDF may continue to depend primarily
on the parameter $S_3\sigma$
even when the second-order Edgeworth approximation begins to break down.
In abstract form, we can write this conjecture as
$$
p[\nu ; n ; \sigma(r_s)] = p[\nu ; S_3(n)\sigma(r_s)] \quad ;
\eqno(\new)
$$
i.e.\ the shape of $p(\nu)$, which could in principle depend on the
spectral slope $n$ and on the fluctuation amplitude $\sigma$
at the smoothing scale $r_s$,
in fact depends on these parameters only through the combination
$S_3(n)\sigma(r_s)$.  In the same notation, we can express the
scaling proposed by Kofman \etal (1993) as
$$
p[\nu ; n ; \sigma(r_s) ] = p[\nu ; \sigma(r_s)].
\eqno(\new)
$$
Figure 8 presents a test of the conjecture (24).
The heavy solid lines in each panel (linear on the left, logarithmic on the
right) show the PDF measured from our $n=0$ simulations at $a=1/2$,
with a smoothing length of 2 cells.  The r.m.s.\ fluctuation on this scale
is $\sigma=0.80$.  The dotted lines, which are mostly obscured by the heavy
solid lines, show the PDF measured from our $n=-1$ simulations at
$a=1/2$, with a smoothing length of 2.82 cells.  The corresponding $\sigma$
is 0.73.  The values of $S_3\sigma$ for these two sets of density fields
are very nearly equal: $3.14\times 0.80 \approx 3.47 \times 0.73 \approx 2.5$.
The light solid curve shows the second-order Edgeworth approximation
for this value of $S_3\sigma$.  We see that the PDFs of the two fields
match very closely, even though neither of them agrees well with the
analytic approximation.  The dashed line shows the PDF of the $a=1/2$,
$n=-1$ simulations at a smoothing length of 2.55 cells, the scale where the
r.m.s.\ fluctuation $\sigma=0.80$ matches that of the $n=0$ density fields.
While this PDF is reasonably close to that of the $n=0$ simulations, it is
clear that the shape of the PDF depends on the spectral slope as well as
on the value of $\sigma$, and that equation (\eqref{2}) offers a more
accurate description of the PDF than equation (\eqref{1}).
Analytically, we can see that equation (\eqref{2}) will hold at the
level of the third-order Edgeworth approximation (equation 14)
if and only if the ratio $S_3^2/S_4$ is independent of $n$.
Bernardeau (in preparation) shows that this ratio is indeed nearly independent
of $n$ for a top hat smoothing filter.

\topinsert
\capskip
\centerline{
\epsfxsize=5.5truein
\epsfbox[45 549 576 738]{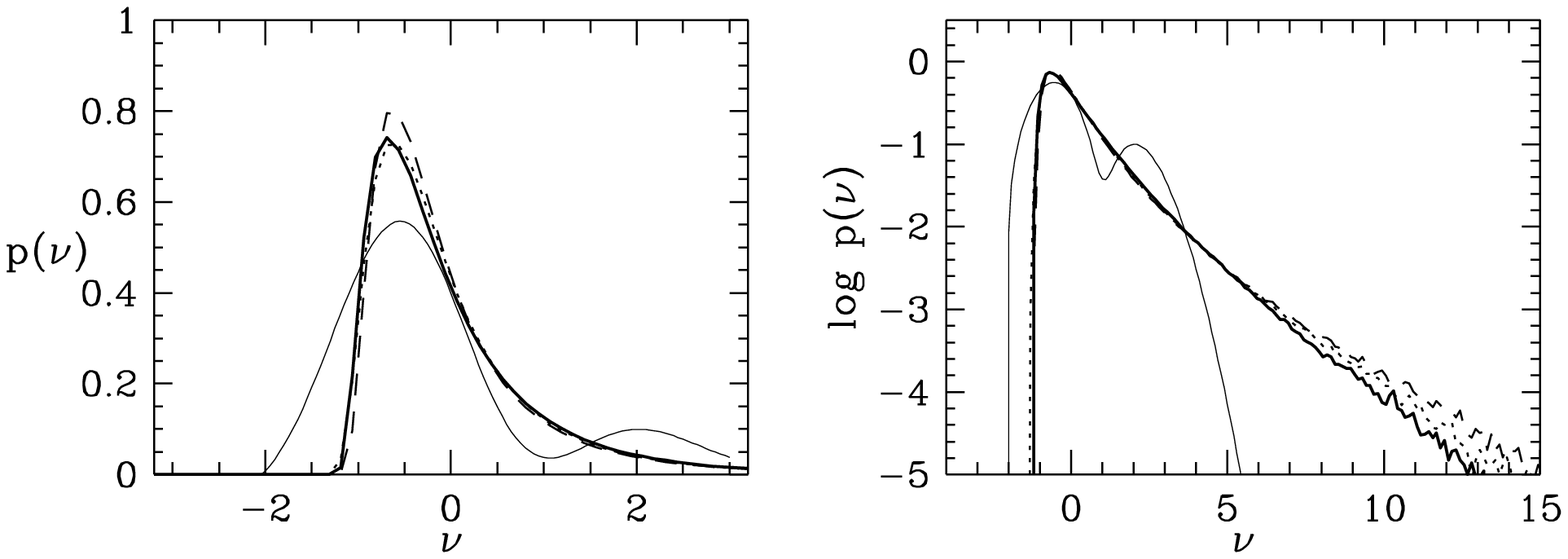}
}
\medskip
{\eightpoint
\noindent {\bf Figure 8} ---
Comparison of density PDFs for different initial power spectra.
The heavy solid lines show the PDF measured from \nbody\ simulations with a
white noise ($n=0$) initial power spectrum.  The r.m.s.\ fluctuation on the
smoothing scale is $\sigma=0.80,$ so $S_3\sigma=3.14\times 0.80\approx 2.5$.
The dotted lines (often obscured by the heavy solid lines)
show the PDF of the $n=-1$ simulations at a scale
where $\sigma=0.73$ and $S_3\sigma=3.47\times 0.73 \approx 2.5$.
Even though the second-order Edgeworth approximation (indicated by the
light solid lines) has failed rather badly, the PDFs of the $n=0$ and
$n=-1$ models are a nearly perfect match on the scales where $S_3\sigma$
is the same.
The dashed lines show the PDF measured from the $n=-1$ \nbody\ simulations
at a scale where $\sigma=0.80$ and $S_3\sigma=3.47\times 0.80 \approx 2.8$.
The shape of the evolved PDF is more nearly a universal function of
$S_3\sigma$ than it is of $\sigma$ alone.
}
\smallskip
\textskip
\endinsert

\bigskip

\noindent
{\bf 5. Biased Galaxy Formation}

Perturbation theory describes the evolution of the mass distribution, but
observations probe the distribution of galaxies.  There are both observational
and theoretical reasons for thinking that galaxies do not evenly trace the
large-scale mass distribution.  On the observational side, we know that
elliptical and spiral galaxies have different clustering properties;
it is clear that the two galaxy types cannot {\it both} trace the mass
independently, and there is no particular reason to expect that the union
of the two classes does trace the mass.  Theoretically, we know that the
collapse epoch of a galaxy scale perturbation will depend on the background
density, so the history of perturbations will vary with environment, and
the efficiency of galaxy formation may vary correspondingly.  Numerical
simulations that include gas dynamics indicate that galaxy formation is
at least somewhat biased towards regions of high background density
(Cen \& Ostriker 1992; Katz, Hernquist \& Weinberg 1992).

For the mass distribution, second-order perturbation theory tells us that
$\dcube=S_3\sigma^4$, if the primordial fluctuations are Gaussian.
Is there a corresponding relation for the galaxy density contrast $\delta_g$?
If we adopt the simplest mathematical relation between the galaxy
and mass density contrasts, the linear bias model $\delta_g=b\delta$,
then the answer is obvious: $\gcube=S_{3g}\sigma_g^4$, where $S_{3g}=S_3/b.$
However, there is no theoretical motivation for the linear bias model, and
while it may be a useful approximation for some purposes, it seems risky
to assume a symmetric bias relation in order to compute a measure of
asymmetry in the probability distribution.  Indeed, one might worry that
allowing a non-linear relation between galaxy density and mass density
would destroy the simple relation between $\dcube$ and $\sigma^4$ predicted
by perturbation theory, but this is not the case, as we shall now see.

Instead of a linear bias, let us adopt the much looser assumption that the
smoothed galaxy density is a non-linear but local function of the smoothed
mass density (see discussion by Coles 1993),
$$
\delta_g=f(\delta).
\eqno(\new)
$$
The second-order Taylor expansion for $\delta_g$ is
$$\eqalignno{
\delta_g &= b\delta + {1\over 2} b_2\delta^2 - {1 \over 2} b_2\sigma^2,
\qquad {\hbox{where}} &(\new{\hbox{a}})\cr
b&=f^{\prime}(0), \qquad b_2=f^{\prime\prime}(0), \qquad \sigma^2=\dsqr.
&(\last{\hbox{b}}) \cr
}
$$
The last term on the right-hand side of equation (\last a)
is required to ensure
that $\langle \delta_g \rangle = 0$.  It is straightforward to use the
expansion (\eqref{1}) to compute $\sigma_g^4$ and $\gcube$ to
$O(\sigma^4)$,
making the substitutions $\dcube=S_3\sigma^4$ and
$\dfour=3\sigma^4 + S_4\sigma^6$ where appropriate.  The result is
$$
\sigma_g^4 = b^4\sigma^4, \qquad
M_{3g} \equiv \gcube = (S_3b^3 + 3b^2b_2)\sigma^4, \eqno(\new)
$$
making the moment ratio for the galaxy distribution
$$
S_{3g} = {M_{3g} \over \sigma_g^4} =
{S_3 \over b} + {3b_2 \over b^2}. \eqno(\new)
$$
For linear bias, $b_2=0$, we recover the earlier result $S_{3g}=S_3/b$.
However, the value of $S_{3g}$ depends sensitively on the shape of
the bias function through the second-derivative term $3b_2/b^2$.
This shape dependence may explain why the value of $S_{3g}$
measured by Bouchet \etal (1993) from
the IRAS redshift survey is rather low, $S_{3g} \sim 1.5$.
IRAS galaxies are underrepresented in rich clusters
relative to optical galaxies, so the ``bias function''
that applies to IRAS galaxies may have negative curvature (negative $b_2$),
pushing $S_{3g}$ below the value of $S_3$ for the mass.
Evidence that cluster-avoidance is an important effect comes
directly from Bouchet {\it et al.}'s (1993) analysis; they find that
double-counting the IRAS galaxies in rich cluster cores, a small
fraction of the total sample, more than doubles the measured
value of $S_{3g}$.

The most important implication of equation (\eqref{2}) is that
$\gcube$ will be proportional to $\sigma_g^4$ on scales where $\sigma_g$ is
small, provided that the linear mass fluctuations are Gaussian and that
the galaxy density is a local function of mass density.
Fry \& Gazta\~naga (1993) have independently derived equation (\eqref{2}),
and they have generalized the result in an important way: by expanding
$\delta_g=f(\delta)$ in higher-order Taylor series, they show that
a local biasing function preserves {\it all} of the moment relations predicted
by perturbation theory (equation 11),
in the limit of small fluctuation amplitude.
One can therefore test the hypotheses of Gaussian
primordial fluctuations and local biasing by examining the moments of the
galaxy count distribution on large scales.

\topinsert
\capskip
\centerline{
\epsfxsize=3.5truein
\epsfbox[116 306 464 728]{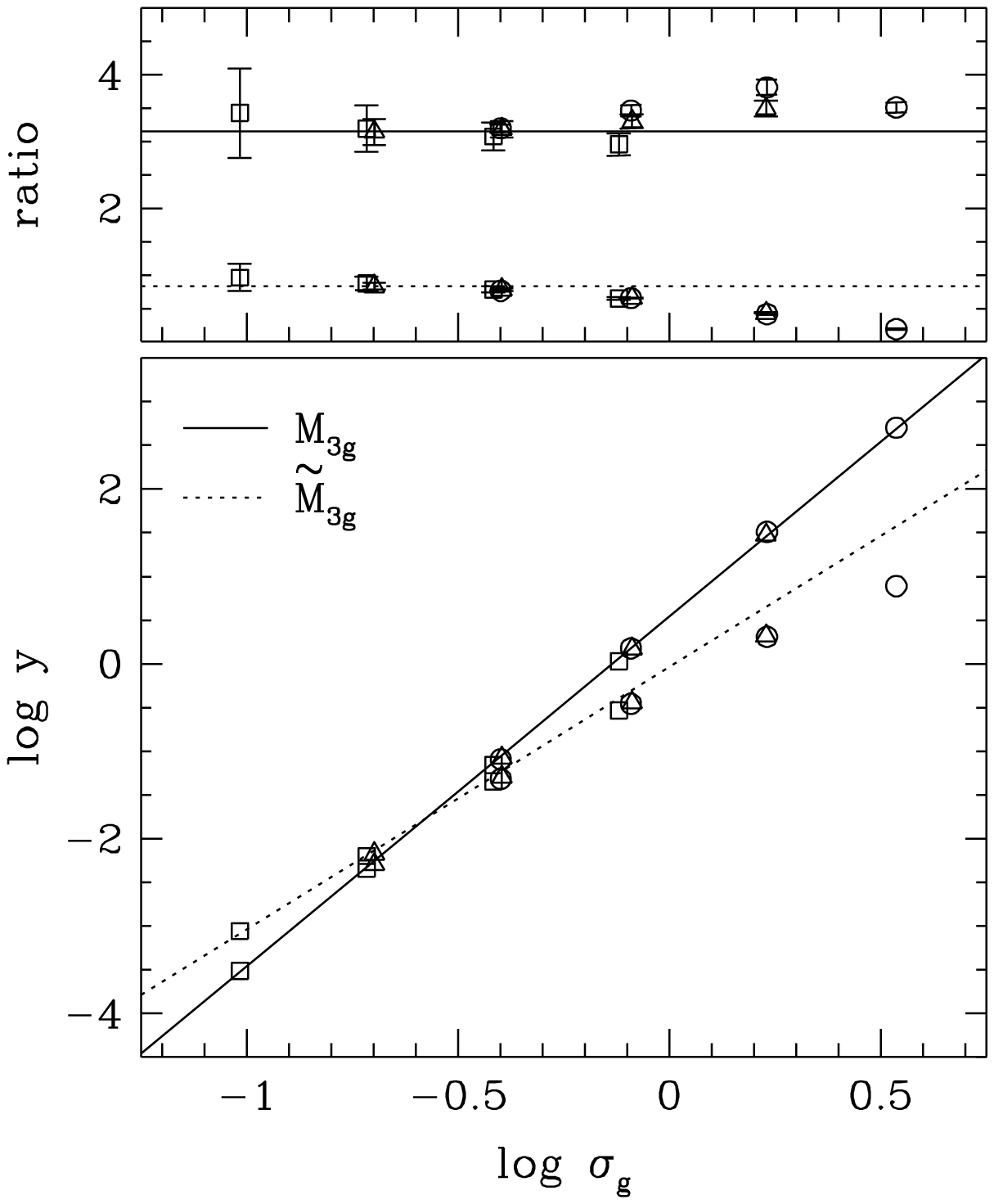}
}
\medskip
{\eightpoint
\noindent {\bf Figure 9} ---
Evolution of $M_{3g}\equiv\gcube$ and
$\tm{3g}\equiv\gg$, where $\delta_g$ is the
contrast of the galaxy density field, related to the mass density
field by the ``biasing function'' (30).
Format is the same as Figure 2.  The solid and dotted lines
show the analytic relations $M_{3g}=S_{3g}\sigma_g^4$ and
$\tm{3g}=\ts{3g}\sigma_g^3$.
Circles, triangles, and squares show results from the biased
\nbody\ density fields at smoothing lengths of 2, 4, and 8 cells, respectively.
The value of $S_{3g}$ is computed from equation (29), so there are no
free parameters to either of these ``fits.''
When $\sigma_g$ is small, biasing preserves the form of the moment relations
predicted by perturbation theory, and with this biasing function,
which is derived from hydrodynamic cosmological simulations,
the perturbative relation $M_{3g} = S_{3g}\sigma_g^4$
remains accurate even at $\sigma_g\approx 3.5$.
}
\medskip
\textskip
\endinsert

We can illustrate these analytic arguments and get a sense of how
they extend into the non-linear regime by considering the bias function
proposed by Cen \& Ostriker (1993; hereafter CO), who incorporate a simple
but physically motivated recipe for galaxy formation into their hydrodynamic
simulations of the cold dark matter model.  CO fit the relation between
galaxy density and mass density in their simulations to a non-linear
functional form,
$$
\log(1+\delta_g) = A + B\log(1+\delta) + C[\log(1+\delta)]^2.
\eqno(\new)
$$
We can apply this transformation directly to the smoothed mass density
fields of our $N$-body simulations and compute the resulting moments.
In their simulations, CO find that the constants $B$ and $C$ depend weakly
on smoothing scale.  We take the values $B=1.5$ and $C=-0.14$ that CO
find for a Gaussian filter scale of $5h^{-1}$ Mpc; at any output time
and smoothing scale, the constant $A$ is then determined by the requirement
that $\langle \delta_g \rangle = 0$.  The details of this biasing procedure
should not be taken too seriously, in part because the resolution of
the CO simulations themselves is only $\sim 0.5 h^{-1}$ Mpc, which is
rather low for inferring rates of galaxy formation.
Nonetheless, this biasing scheme reflects the sort of relation between
galaxy and mass density that one might expect in a rather broad class
of theoretical scenarios.
(The simulations of Katz \etal [1992] and Evrard, Summers \& Davis [1993]
have much higher spatial resolution in
galaxy forming regions, but in each case the simulation cube is
$\sim 10\hmpc$, too small
a volume in which to measure a meaningful biasing function.)

\topinsert
\capskip
\centerline{
\epsfxsize=3.5truein
\epsfbox[116 306 464 728]{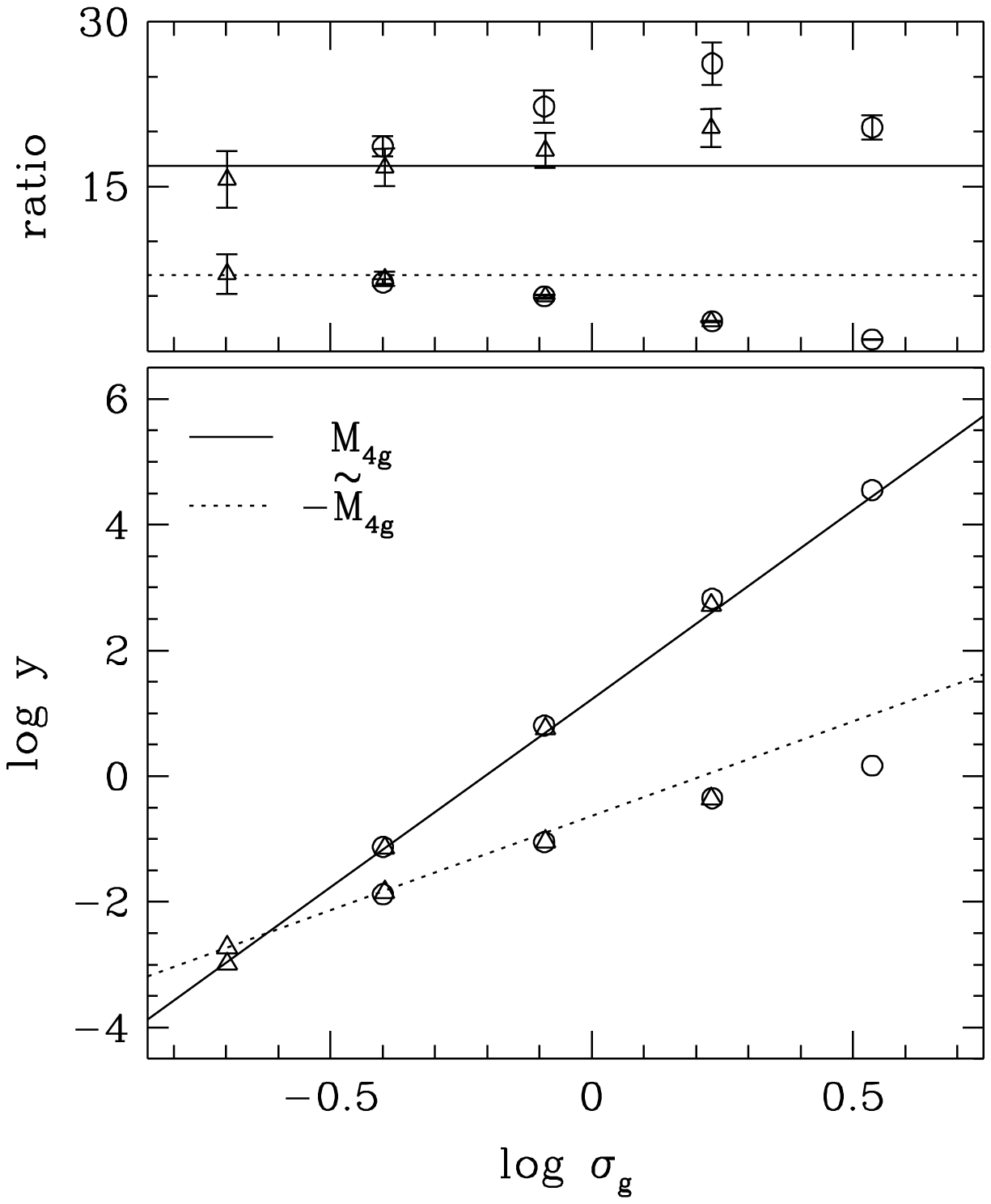}
}
\medskip
{\eightpoint
\noindent {\bf Figure 10} ---
Evolution of $M_{4g}$ and $-\tm{4g}$, in the same format as
Figure 3.  Solid and dotted lines in the lower panel show the
analytic predictions $M_{4g}=S_{4g}\sigma_g^6$ and $-\tm{4g}=-\ts{4g}\sigma^3$.
Circles and triangles show results from
the biased \nbody\ density fields at smoothing lengths of 2 and 4 cells,
respectively.  The value of $S_{4g}$ is computed analytically using the
value of $S_4$ measured for the simulation mass density fields,
so there are no free parameters to these ``fits.''
In the upper panel, the values of $-\tm{4g}/\sigma_g^3$ (bottom set of points
and horizontal dotted line) are multiplied by a factor of 30, to make
them easily visible on this plot.
}
\medskip
\textskip
\endinsert

Figure 9 plots $M_{3g}$ and $\tm{3g}$ against $\sigma_g$ for our $N$-body
simulations, where $\delta_g$ and $\delta$ are related by equation (\eqref{1}).
The format is the same as Figure 2.
The solid and dotted lines in the lower panel
show the perturbation theory predictions,
$M_{3g}=S_{3g}\sigma_g^4$ and
$\tm{3g}=\ts{3g}\sigma_g^3$, respectively.
The upper panel plots the ratios $M_{3g}/\sigma_g^4$ and
$\tm{3g}/\sigma_g^3$ for the simulations and the analytic approximations.
We compute $S_{3g}=3.15$
from equations (\eqref{2}) and (\eqref{1}) in the limit of
$\dsqr \rightarrow 0$; in this limit, the constant $A$ in the biasing
function (\last) goes to zero.  The results are strongly reminiscent of
those for the mass distribution, shown in Figure 2.
When $\sigma_g$ is small, the $N$-body points sit precisely
on the perturbation theory lines, as expected.  As $\sigma_g$ approaches unity,
the analytic approximation for $\tm{3g}$ begins to fail,
but $M_{3g}$ follows the
perturbation theory prediction remarkably closely
up to the last point computed, where $\sigma_g \approx 3.5$.
This continuing agreement is not guaranteed by equation (\eqref{3}),
since the derivation of that equation is based on a second-order Taylor
expansion which is no longer valid when $\sigma_g \simgt 1$.

Figure 10 plots $M_{4g}$ and $\tm{4g}$ against $\sigma_g$,
in the same format as Figure 3.
The solid line in the lower panel
shows the perturbative relation $M_{4g}=S_{4g}\sigma_g^6$.
We compute $S_{4g}=16.9$ using equation (10) of Fry \& Gazta\~naga (1993),
assuming $S_4=20$ for the mass (from Figure 3).
The dotted line shows the Edgeworth approximation $\tm{4g}=\ts{4g}\sigma_g^3$,
with $\ts{4g}$ computed from equation (21).
The upper panel displays the ratios $M_{4g}/\sigma_g^6$ and
$\tm{4g}/\sigma_g^3$, with the latter multiplied by a factor of 30 for
visibility.
The analytic and numerical results agree when $\sigma_g$ is low, as they
should.  The Edgeworth approximation of $\tm{4g}$ overestimates
the \nbody\ value at larger $\sigma_g$, but the value of $M_{4g}$
stays close to the relation predicted by perturbation theory up to
the last point, $\sigma_g \approx 3.5$.
We have also compared the full PDFs of the biased density fields to
those computed from the Edgeworth expansion.  The results are similar to
those shown in Figure 4 for the mass distribution: good agreement for
$\sigma_g \simlt 1/2$, and poor agreement beyond.

If the primordial mass fluctuations are non-Gaussian, e.g.\ if they have
intrinsic skewness or kurtosis, then the linear term of the biasing
function will transfer these intrinsic moments to the galaxy fluctuations,
and moments of the galaxy counts will not obey the hierarchical relations
of equation (11).  Galaxy counts may also violate equation (11) if the galaxy
density is not tightly coupled to the local mass density.  One could imagine,
for instance, that the galaxy density obeys $\delta_g=f(\delta)$ on
average but with substantial scatter about the mean trend.  Scatter might
arise if the efficiency of galaxy formation is sensitive to the pressure
of the local intergalactic medium or to ionizing radiation from nearby quasars.
In the limit where scatter about the mean relation overwhelms the
trend predicted by the mean relation itself, it is clear that moments of
the galaxy distribution will reflect the moments of the ``scatter function''
rather than moments of the underlying mass distribution, and there is
no reason to expect these moments to obey the special relations implied
by equation (11).  For example, Frieman \& Gazta\~naga (1993) show
that the ``cooperative galaxy formation'' scheme, proposed by
Bower \etal (1992) as a possible way to reconcile the standard CDM
model with the galaxy angular correlation function of the APM
survey (Maddox \etal 1990), predicts a strong shift in the
relation between skewness and variance of galaxy counts at scales
$\sim 10-20\hmpc$.
Precise observational confirmation of hierarchical
relations between moments of galaxy counts would provide evidence in
favor of Gaussian primordial fluctuations {\it and} important constraints
on the process of galaxy formation itself.

\bigskip

\noindent
{\bf 6. Discussion \hfill}

The comparisons in Section 4 provide encouraging news both for
the perturbative analytic approach and for the \nbody\ simulations themselves.
Most previous tests of cosmological \nbody\ methods have examined either
the linear growth of fluctuations or strongly non-linear problems like
pancake collapse.  When the variance is small, the non-linear effects
discussed in this paper are quite subtle, as evidenced by our plots of
dimensionless quantities extending to $\sim 10^{-4}$ and even below.
Nonetheless, the moments of our \nbody\ density fields match the perturbative
calculations perfectly when the variance is small.
These precise measures test the \nbody\ method in a new regime, that of
weakly non-linear clustering, and the agreement with analytic theory
strengthens our faith in the reliability of the simulations.
Most cosmological \nbody\ studies use a cubic lattice, the Zel'dovich
approximation, and periodic boundaries to set up initial conditions, just
as we do.  The match to second- and third-order
perturbation theory in the weakly non-linear regime
is significant because it shows that such initial conditions allow
\nbody\ simulations to settle into
the correct non-linear solution for the evolution of the
mass distribution.  Success is not guaranteed by the initial conditions
algorithm itself, since the Zel'dovich approximation does {\it not}
yield the correct relations between moments of the density field
(Grinstein \& Wise 1987; JBC; BJDB; see discussion in \S 4.1).

The \nbody\ simulations confirm the correctness of the analytic
calculations, including the moment calculations of JBC and BJDB and
our new results for PDFs, $\tm3$,
and $\tm4$ of the smoothed density field and the smoothed velocity divergence.
The simulations
show that these results continue to hold on large scales even
when small-scale clustering is strongly non-linear.  This conclusion
is unsurprising; more remarkable is the fact that the perturbative
approximations do not break down rapidly in the non-linear regime.
In particular, the skewness
and kurtosis of the density field stay impressively close to the
predictions of perturbation theory even when $\sigma=2$.
The second-order Edgeworth approximation to the PDF of the density
or velocity divergence remains accurate until $|S_3\sigma|$,
the magnitude of the skewness, reaches one, and the third-order approximation
remains accurate slightly longer.  Calculations of
$\tm3$ and $\tm{3\theta}$ based on the Edgeworth series
match the \nbody\ results to $\sim 15\%$ when
$\sigma \sim 1/2$ and $\sim 30\%$ when $\sigma \sim 1$.
We have applied our techniques specifically to the case of Gaussian
initial conditions.  We believe that a similar treatment is possible
for non-Gaussian initial conditions, though the method requires
some modification because a low-order Edgeworth expansion may not
provide a good description of the linear theory PDF in a non-Gaussian model.

One of the encouraging results of this paper (derived independently by
Fry \& Gazta\~naga 1993) is that ``biased'' galaxy formation preserves
the relation between the skewness and variance of the density field
predicted by perturbation theory,
provided that the galaxy density is a local function of the mass density.
This fact can be demonstrated analytically in the limit of small fluctuations,
and once again our tests on numerical simulations show that the
predictions of perturbation theory continue to hold remarkably well in
the fully non-linear regime, at least for the biasing relation proposed
by Cen \& Ostriker (1993).  By studying galaxy counts on large scales,
one can learn about both the nature of primordial fluctuations and the
physics of galaxy formation.

We have limited
the analysis in this paper to the case of $\Omega=1$.
Perturbation theory predicts that the $S_3$ coefficient for the density
field should have only a very weak dependence on $\Omega$ (BJCP).
However, for the velocity divergence there is a fairly strong dependence,
roughly $S_{3\theta} \propto \Omega^{-0.6}$ (BJDB),
so predictions for the moment relations (Figures 5 and 6)
and the PDF (Figure 7) depend significantly on $\Omega.$
If perturbation theory works equally well for low-$\Omega$ models,
and we have every reason to think that it will, then the moments and
the PDF of the velocity divergence can be used to constrain the density
parameter provided that (a) one adopts the hypotheses of gravitational
instability and Gaussian fluctuations, and (b) one can obtain a reliable
estimate of the velocity field over a sufficiently large volume.
Since this technique does not use the galaxy density field, it is independent
of biased galaxy formation so long as galaxies provide fair tracers of
the large-scale velocity field.
We address these ideas more fully elsewhere (BJDB).

A close relative of the velocity divergence technique mentioned
above is the ``reconstruction'' method of Nusser \& Dekel (1993,
hereafter ND), who attempt to recover the PDF of the initial
density fluctuations from the divergence of the present day
velocity field by applying the Zel'dovich approximation.
They find that the velocity field inferred by POTENT
(Bertschinger \& Dekel 1989; Bertschinger \etal 1990) is consistent
with Gaussian initial conditions if $\Omega=1$, but not if
$\Omega=0.3$.  We have two cautionary remarks to make about this
conclusion (in addition to the caveats listed by ND themselves).
First, as discussed in \S 4.1 and \S 4.3, the Zel'dovich approximation
makes large errors in predicting the skewness of the velocity
divergence, which is the simplest measure of asymmetry in its PDF.
Second, the residuals between ND's recovered initial PDF and a
true Gaussian, plotted in figure 5 of ND, bear a remarkable
resemblance in shape to the Gaussian derivatives $\phi^{(4)}$
and $\phi^{(6)}$ (see Figure 1).
Since these derivatives appear in the Edgeworth approximation to
the evolved PDF, their appearance in ND's residuals may
indicate a systematic dynamical failure of their reconstruction
method.  However, the magnitude of such a failure is constrained
by the method's reasonable success when applied to \nbody\
simulations with known initial conditions
(ND; Gramman, Weinberg \& Nusser, in preparation).

With the rapid growth in galaxy redshift and peculiar velocity data,
the regime of weakly non-linear clustering is becoming increasingly
accessible to observations.  By comparing PDFs and moments of the density
and velocity divergence fields to the predictions discussed here,
we can test the hypothesis that structure in the universe formed by
gravitational instability from Gaussian primordial fluctuations.
In the analysis of galaxy density fields
one must introduce assumptions about the relation between galaxies and mass,
and in the analysis of velocity fields one must introduce assumptions
about $\Omega$, but by examining structure over a variety of scales
one can check all of the input assumptions for internal consistency.
Application of these methods to high quality, large volume
data sets should therefore teach us a great deal about the formation
of galaxies and the origin of large-scale structure.

\bigskip
\bigskip

We are very grateful to Changbom Park for the use of his PM \nbody\ code
and to Jeremiah Ostriker for discussions that led to \S 5 of this paper.
We acknowledge helpful discussions with Francis Bernardeau, Stephan
Colombi, Mirt Gramman, Lev Kofman, Adi Nusser, and Michael Strauss.
RJ, PA, and MC acknowledge support from Polish Government
Committee for Scientific Research (KBN) grant number 2.1243-91.01.
DHW acknowledges a fellowship from the W. M. Keck Foundation
and additional support from the Ambrose Monell Foundation and
NSF grant PHY92-45317.

\vfill\eject

\noindent
{\bf Appendix \hfill}

Our purpose here is to derive $\dd$, using gravitational instability
theory, and the assumption that $\delta_1$ is a Gaussian random field.
Since $\delta_2 = O(\delta_1^2)$, we can use the expansion
$$
|\delta| \equiv \sqrt{(\delta_1 +\delta_2 + \ldots)^2} =
|\delta_1|\,(1 + \delta_2\,/\delta_1) + O(\delta_1^3)
\eqno(A1)
$$
This gives
$$
\delta |\delta| = \delta_1 |\delta_1| +
2\,\delta_2\,|\delta_1| + O(\delta_1^4)\; .
\eqno(A2)
$$
By symmetry, the mean value of the first term above is zero,
and the lowest order contribution to $\dd$
comes from the second term. To calculate
$$
\dd =  2\,\langle \delta_2 \,|\delta_1|\rangle \; ,
\eqno(A3)
$$
we need the joint probability distribution for $\delta_1$ and
$\delta_2$. For density fluctuations in an expanding universe,
filled with a pressureless non-relativistic fluid, the relevant
perturbative solutions are given by (Juszkiewicz \& Bouchet 1992)
$$
\delta_1 = D(t)\e({\bf x}) \; ,
\eqno(A4)
$$
$$
\delta_2 = D^2(t)\,[{\textstyle {2\over 3}}(1 + \k)\e^2
+ \nabla \e \cdot \nabla \Phi + ({\textstyle {1\over 2}} - \k)
\tau_{\alpha\beta}\tau_{\alpha\beta}] \; ,
\eqno(A5)
$$
where we use the Einstein summation convention,
$t$ is the cosmological time, $\;{\bf x} = \{ x_{\alpha} \},
\; \alpha = 1,2,3,\;$ are comoving coordinates, and
$D(t)$ is the linear order fluctuation growth rate (cf. Peebles 1980).
Without loss of generality, in our remaining calculations below we
will always assume $D(t) = 1$.
The quantities $\Phi$ and $\tau_{\alpha\beta}$ are proportional
to the linear order Newtonian potential and the shear tensor, respectively:
$$
\nabla^2 \Phi \equiv \e({\bf x}) \; ;\quad \tau_{\alpha\beta}
\equiv ({\textstyle {1\over 3}}\delta_{\alpha\beta}\nabla^2 - \nabla_\alpha
\nabla_\beta)\Phi({\bf x}) \; .
\eqno(A6)
$$
Here $\delta_{\alpha\beta}$
is the Kronecker delta. The parameter $\k$ is
a slowly varying function of cosmological time. For densities in
the range $0.05\le \Omega \le 3$, it is well approximated by
$\k[\Omega(t)] \approx (3/14)\,\Omega^{-2/63}$ (BJCP).
It is convenient to introduce the Fourier transform
$$
\e({\bf x}) = \int \; {d^3k\over(2\pi)^{3/2}}\; e^{-i{\bf k\cdot x}}\,
\e_{\bf k} \; .
\eqno(A7)
$$
The {\it power spectrum} is defined by the relation
$$
\langle \e_{\bf k} \, \e_{\bf k'}\rangle = P(k)\;\delta_D({\bf k + k'})\; ,
\eqno(A8)
$$
where $\delta_D$ is the Dirac delta function. The Fourier transform
of $\Phi$ is $\;\Phi_{\bf k} = -\e_{\bf k}/k^2$. The transforms of
the potential and density gradients, $\;\nabla \Phi\;$ and
$\;\nabla\e \;$, are $\;i{\bf k}\,\e_{\bf k}/k^2\;$ and
$\; -i{\bf k}\,
\e_{\bf k}\;$, respectively. The shear tensor can be represented
by the Fourier integral
$$
\tau_{\alpha\beta}({\bf x}) = \int\;{d^3k\over(2\pi)^{3/2}}\;
({\textstyle {1\over 3}}\delta_{\alpha\beta} - \hat k_{\alpha}
\hat k_{\beta})\; \e_{\bf k} \; e^{-i{\bf k\cdot x}} \; ,
\eqno(A9)
$$
where $\hat k_{\alpha} \equiv k_{\alpha}/k$. To calculate $\dd$,
we need the joint probability density,
$p(\e, \nabla\e, \nabla\Phi, \tau_{\alpha\beta})$.
Since $\e$ is Gaussian, $p(\e, \ldots)$
is a multivariate Gaussian distribution,
entirely determined by its covariance matrix.
There is no need to calculate this horrifying 12 x 12 matrix
explicitly because $\e$ is statistically independent
from the remaining variables. Indeed, $\e$ is
correlated only with itself:
$$
\la\e^2({\bf x})\ra = \int\;{d^3k\,d^3k'\over(2\pi)^3}\la\e_{\bf k}\,
\e_{\bf k'}\ra \;e^{-i{\bf( k + k')\cdot x}} = \int \; {d^3k\over(2\pi)^3}
\;P(k) \equiv \sigma^2 \; ,
\eqno(A10)
$$
while
$$
\la\e \,\nabla\e\ra = \int \; {d^3k\over(2\pi)^3}\;i{\bf k}\,P(k) = 0 \; ,
\qquad \la\e\,\nabla\Phi\ra = \int\;{d^3k\over(2\pi)^3}\;{i{\bf k}\over k^2}
\,P(k) = 0 \; ,
\eqno(A11)
$$
and
$$
\la\e\,\tau_{\alpha\beta}\ra = 0 \; .
\eqno(A12)
$$
To derive the last expression above,
we used equations (A9) and (A7),
as well as the identity
$$
\int P(k)\,\hat k_\alpha\,\hat k_{\beta}\;{d^3k\over(2\pi)^3} \; = \;
{\textstyle {1\over 3}}\,\sigma^2\,\delta_{\alpha\beta} \; ,
\eqno(A13)
$$
which can be verified easily
by integrating over the appropriate angles.
Equations (A11) and (A12) imply
$$
p(\e,\nabla\e,\nabla\Phi,\tau_{\alpha\beta})
= p(\e)\;p(\nabla\e,\nabla\Phi,\tau_{\alpha\beta}) \; .
\eqno(A14)
$$
Now, let us represent
$\delta_2$ in equation (A5) as a sum of two terms,
$$
I_1 \equiv {\textstyle{2\over 3}}\;(1 + \k)\;\e^2 \, , \qquad
{\rm and} \qquad I_2 \equiv \nabla\e \cdot \nabla \Phi +
({\textstyle{1\over 2}} - \k)\;\tau_{\alpha\beta}\,
\tau_{\alpha\beta} \; .
\eqno(A15)
$$
The factorization of the joint probability density
(eq.[A14]) allows us to write
$$
\la |\e|\,I_2\ra = \la |\e|\ra\,\la I_2 \ra \; .
\eqno(A16)
$$
The second and higher order corrections to $\delta_1$ must preserve
$\la \delta\ra = 0$ because mass is conserved (cf. \S 18 in Peebles
1980). This implies $\la \delta_2 \ra \equiv \la I_1 + I_2 \ra = 0$,
and therefore
$$
\la I_2 \ra = - \la I_1 \ra  =
-{\textstyle {2\over 3}}\,(1 + \kappa)\,\sigma^2 \; .
\eqno(A17)
$$
Now the calculation of $\dd$ is reduced to
$$
\dd = 2\,\la |\e|\,(I_1 + I_2)\ra = 2\,\la |\e|\,I_1\ra
- 2\,\la |\e|\ra\,\la I_1\ra \; ,
\eqno(A18)
$$
and all averages above involve only the marginal distribution
$p(\e) = \exp(-\e^2/2\sigma^2)
\;/\sqrt{2\pi\sigma^2}$. The contribution from $I_1$ amounts to
$$
2\, \la |\e|\,I_1\ra =
{8(1+\kappa)\over 3\sigma \sqrt{2\pi}}\,
\int_0^{\infty}\;\e^3\;e^{-\e^2/2\sigma^2}\;d\e
= \sqrt{128\over 9\pi}\;(1 + \k)\;\sigma^3 \; .
\eqno(A19)
$$
The mean value of $|\e|$ equals
$$
\la |\e| \ra = {2\over\sigma\,\sqrt{2\pi}}\,\int_0^\infty\;
\e\, e^{-\e^2\,/2\sigma^2}\,d\e = \sqrt{2\over \pi}\,\sigma \; ,
\eqno(A20)
$$
so the contribution from $I_2$ is
$$
- 2\,\la |\e|\ra\,\la I_1\ra = - \sqrt{32\over 9\pi}\,(1 + \kappa)\sigma^3\; .
\eqno(A21)
$$
Combining this with eq.(A19), we obtain
$$
\dd = \sqrt{32\over 9\pi}\;(1 + \k)\;\sigma^3 = \sqrt{2\over 9\pi}
\;S_3\;\sigma^3 \; ,
\eqno(A22)
$$
where we use the expression
$$
S_3 = 4\,(1 + \k) \; ,
\eqno(A23)
$$
obtained for the unsmoothed field by BJCP. Our above
result is in agreement with equation (17) in the main text. However,
equation (17) is more general than our present result, since its
validity is not restricted to the unsmoothed field and its particular
value of $S_3$. It is valid for arbitrary filters and arbitrary power
spectra. An attempt to include smoothing would make the calculation
``from first principles'', like the one we just finished, much more
complicated. Instead of vanishing one-point moments, like
$\la\nabla\e({\bf x})\,\e({\bf x})\ra = 0$, we would
have to deal with correlation
functions, like $\la\nabla\e({\bf x + r})\,\e({\bf x})\ra$, which
generally do not vanish for $r \ne 0$. As a result, we would not be
able to reduce the dimensionality of the PDF through factorization
formulae like equation (A14). In such cases it is much simpler
to do as we did in section 3: use the Edgeworth expansion,
with the values of cumulants (like $S_3$)
calculated from perturbation theory for the filtered field.

\vfill\eject

\def\ref{ \par \hangindent=20pt \hangafter=1 \noindent }
\parskip=6pt

\centerline{\bf REFERENCES}

\ref
Abramowitz, M. \& Stegun, I. A. 1964, Handbook of Mathematical Functions
(Washington: National Bureau of Standarts)

\ref
Allen, T. J., Grinstein, B. \& Wise, M. B. 1987, Phys. Lett. B, 197, 66

\ref
Bardeen, J. M., Steinhardt, P. J. \& Turner, M. S. 1983  Phys. Rev. D, 28, 679

\ref
Barriola, M. \& Vilenkin, A. 1989, Phys. Rev. Lett., 63, 341

\ref
Bernardeau, F. 1992, ApJ, 392, 1



\ref
Bertschinger, E. \& Dekel, A. 1989, ApJ, 336, L5

\ref
Bertschinger, E., Dekel, A., Faber, S.M., Dressler, A. \& Burstein, D. 1990,
ApJ, 364, 370

\ref
Bouchet, F.R., Adam, J.C. \& Pellat, R. 1985, A\&A, 144, 413

\ref
Bouchet, F. R., Juszkiewicz, R., Colombi, S., \& Pellat, R. 1992a, ApJ, 394, L5
(BJCP)

\ref
Bouchet, F. R., Davis, M., \& Strauss, M. A. 1992b, in
The Distribution of Matter in the Universe, ed. G. A. Mamon \& D. Gerbal
(Meudon: Observatoire de Paris), 287

\ref
Bouchet, F. R., Strauss, M. A., Davis, M., Fisher, K. B., Yahil, A., \& Huchra,
J.P. 1993, ApJ, in press

\ref
Bower, R.G., Coles, P., Frenk, C.S. \& White, S.D.M. 1993, ApJ, 405, 403

\ref
Cen, R., \& Ostriker, J. P. 1992, ApJ, 399, L113

\ref
Cen, R., \& Ostriker, J. P. 1993, ApJ, submitted

\ref
Coles, P. 1993, MNRAS, 262, 1065

\ref
Coles, P., \& Frenk, C. 1992, MNRAS, 253, 727

\ref
Cram\'er, H. 1946, Mathematical Methods of Statistics
(Princeton: Princeton Univ. Press)

\ref
Efstathiou, G., Frenk, C. S., White, S. D. M., \& Davis, M. 1988, MNRAS, 235,
715

\ref
Evrard, A.E., Summers, F.J. \& Davis, M. 1993, ApJ, in press

\ref
Frieman, J. \& Gazta\~naga, E. 1993, ApJ, submitted

\ref
Fry, J. N. 1984, ApJ, 279, 499

\ref
Fry, J. N. \& Gazta\~naga, E. 1993, ApJ, submitted

\ref
Gerhard, O. E. 1993, MNRAS, in press

\ref
Goroff, M. H., Grinstein, B., Rey, S.-J., \& Wise, M. B. 1986, ApJ, 311, 6
(GGRW)

\ref
Grinstein, B., \& Wise, M. B. 1987, ApJ, 320, 448

\ref
Guth, A. H., \& Pi, S.-Y. 1982, Phys. Rev. Lett. 49, 1110

\ref
Hawking, S. W. 1982, Phys. Lett. B, 115, 295

\ref
Juszkiewicz, R., \& Bouchet, F. R. 1992, in The Distribution of Matter in the
Universe, ed. G. A. Mamon \& D. Gerbal (Meudon: Observatoire de Paris), 301

\ref
Juszkiewicz, R., Bouchet, F. R., \& Colombi, S. 1993, ApJ, in press (JBC)

\ref
Katz, N., Hernquist, L., \& Weinberg, D. H. 1992, ApJ, 399, L109

\ref
Kofman, L., \& Pogosyan, D. Yu. 1988, Phys. Lett. B, 214, 508

\ref
Kofman, L., Bertschinger, E., Gelb, J. M., Nusser, A., \& Dekel, A. 1993,
preprint

\ref
Little, B., Weinberg, D.H. \& Park, C. 1991, MNRAS, 253, 295

\ref
Lucchin, F., Matarrese, S., Melott, A. L., \& Moscardini, L. 1993, ApJ, in
press

\ref
Maddox, S.J., Efstathiou, G., Sutherland, W.J. \& Loveday, J. 1990, MNRAS,
242, 43P

\ref
Melott, A.L. 1986, Phys. Rev. Lett., 56, 1992

\ref
Melott, A.L., Weinberg, D.H. \& Gott, J.R. 1988, ApJ, 328, 50

\ref
Moutarde, F., Alimi, J.-M., Bouchet, F.R., Pellat, R. \& Ramani, R. 1991,
ApJ, 382, 377

\ref
Nusser, A., \& Dekel, A. 1993, ApJ, 405, 437 (ND)

\ref
Padmanabhan, T. \& Subramanian, K. 1993, ApJ, 410, 482

\ref
Park, C. 1990, PhD thesis, Princeton University

\ref
Park, C. 1991, ApJ, 382, L59

\ref
Park, C., \& Gott, J. R. 1991, MNRAS, 249, 288

\ref
Peebles, P. J. E. 1980, The Large-Scale Structure of the Universe
(Princeton: Princeton Univ. Press)

\ref
Salopek, D. S., Bond, J. R., \& Bardeen, J. M. 1989, Phys. Rev. D, 40, 6

\ref
Silk, J., \& Juszkiewicz, R. 1991, Nature, 353, 386

\ref
Starobinsky, A. A. 1982, Phys. Lett. B, 117, 175

\ref
Turok, N. 1989 Phys. Rev. Lett., 63, 262

\ref
van der Marel, R.P., \& Franx, M. 1993, ApJ, 407, 525

\ref
Zel'dovich, Ya. B. 1980, MNRAS, 192, 663


\bye